\documentclass[onecolumn,12pt,draftcls]{IEEEtran}
\usepackage{amsmath,epsfig,amssymb}
\usepackage{color}
\usepackage{mysymbol}
\usepackage{afterpage}
\usepackage[normalem]{ulem}

\newtheorem{remark}{Remark}

\def \bW {\bbW}
\def \cV {\ccalB}
\def \bx {\bbx}

\def \bH {\bbH}
\def \bI {\bbI}
\def \E {{\mathbb{E}}}

\def \cB {\ccalB}

\def \cut {{\mathrm{cut}}}
\def \rcut {{\mathrm{rcut}}}

\def \tr {{\mathrm{tr}}}
\def \lambmax {{\lambda_\mathrm{max}}}

\title{Backhaul-Constrained Multi-Cell Cooperation Leveraging Sparsity and Spectral Clustering}

\author{Swayambhoo Jain, Seung-Jun Kim and Georgios B. Giannakis%
\thanks{Manucript revised May 29, 2015; accepted September 14, 2015.}%
\thanks{S. Jain and G. B. Giannakis are with the Department of Electrical and Computer Engineering, University of Minnesota, Minneapolis, MN 55455, USA. E-mail: \{{\tt jainx174,georgios}\}{\tt @umn.edu}. Seung-Jun Kim is with the Department of Computer Science and Electrical Engineering, University of Maryland, Baltimore County, 1000 Hilltop Circle, Baltimore, MD 21250, USA. E-mail: {\tt sjkim@umbc.edu}. }%
\thanks{Part of this work was presented at the IEEE 13th Workshop on Signal Processing Advances in Wireless Communications (SPAWC) held in Cesme, Turkey during June 17th--20th, 2012.}
\thanks{This work was supported in part by by NSF grants 1247885, 1343248, 1423316, 1442686, 1508993, 1509040; and
	ARO grant W911NF-15-1-0492.}%
}

\begin{document}

\maketitle

\begin{abstract}
Multi-cell cooperative processing with limited backhaul traffic is studied for cellular uplinks. Aiming at reduced backhaul overhead, a sparse multi-cell linear receive-filter design problem is formulated. Both unstructured distributed cooperation as well as clustered cooperation, in which base station groups are formed for tight cooperation, are considered. Dynamic clustered cooperation, where the sparse equalizer and the cooperation clusters are jointly determined, is solved via alternating minimization based on spectral clustering and group-sparse regression. Furthermore, decentralized implementations of both unstructured and clustered cooperation schemes are developed for scalability, robustness and computational efficiency. Extensive numerical tests verify the efficacy of the proposed methods.
\end{abstract}

\section{Introduction}
\label{sec:intro}
\subsection{Background}
Rapid growth in wireless data traffic due to wide adoption of mobile devices has reinforced the research and development efforts to boost spectral efficiency of cellular networks. As the cell deployment becomes denser with a smaller cell size, it is recognized that inter-cell interference is a major bottleneck in provisioning desired link quality uniformly in the service area. Multi-cell cooperative processing (MCP), also referred to as network multi-input multi-output (MIMO), has shown promises for alleviating, and in many cases, exploiting, inter-cell interference. The idea is to have the base stations (BSs) collaborate in transmission and reception to/from the mobile stations (MSs), with the necessary coordination among the BSs taking place through the backhaul links~\cite{KFV06}. The premise is that heavy interference experienced by the terminals residing at the cell boundaries is mitigated such that overall network throughput as well as fairness among users are improved. Field trials have verified that significant gain is indeed realizable~\cite{MGF11,IDM11}.

There are a number of approaches for implementing MCP~\cite{GHH10,IDM11}. The BSs can loosely cooperate to avoid strong interference to/from the cell-edge users by exchanging channel state and scheduling information, and performing judicious power control and spatial filtering. When a high-speed low-latency backhaul network is available, more proactive and tighter coordination can be employed, where the antennas located at different BSs are transformed into a virtual antenna array. In the downlink, collaborating BSs can share transmitted symbols to engage in cooperative MIMO transmission. In the uplink, received signal samples at different BSs can be processed together to jointly decode multiple MSs. Such cooperative transmission and reception allow exploiting inter-cell interference constructively to actually improve link quality. Sophisticated coding strategies may also be adopted to expand the achievable rate region. As the dominant computational burden is still placed on the BSs, MSs can enjoy performance improvement at minimal increase in complexity.


To achieve the promised gain of MCP, important practical issues such as the finite backhaul capacity, latency, and synchronization issues must be addressed~\cite{yang2013we}. Unlimited backhaul capacity assumed in many of the earlier theoretical works is impractical for large cellular systems~\cite{HaW93,ShZ01}. Rather, the backhaul traffic volume must be contained in practice. In the same token, it may be infeasible to connect a large number of BSs to a central coordinator due to limitations in long-range backhauling. Instead, more localized communication structure may be desirable~\cite{AEH08TWC,SSP09TIT}. Also, the channel state information may contain uncertainties due to practical training, quantization, and latency~\cite{HKD11TSP}. Maintaining synchronization across multiple cells is challenging. Moreover, complexity of various MCP algorithms may become prohibitive as the size of the network grows, calling for scalable distributed solutions.

The main focus of this work is to incorporate the limited backhaul rates. For instance, in the recent 3GPP (3rd Generation Partnership Project) Long Term Evolution (LTE) and LTE-Advanced cellular standards, the BSs can coordinate their trnasmission and reception by communicating over the logical X2 interface, or the S1 interface (through the core network), in the Coordinated Multi-Point (CoMP) transmission/reception mode~\cite{3GPP_TR36.814}. The backhaul rate requirements vary depending on the specific methods of coordination. Coordinated scheduling of multiple BSs in the sub-frame timescale requires backhauling on the order of several hundred kbps~\cite{Brueck10}. For joint decoding, the requirement goes up to several Mbps~\cite{MGF11,IDM11,yang2013we}. Since analog signal samples (albeit in quantized or compressed formats) received from distributed antennas must be pooled together for joint processing, the backhaul overhead is quite significant in the uplink, compared to the downlink case where digital signals need to be shared~\cite{HFG09}. Our focus in this work is on the most backhaul-intensive uplink joint decoding scenario.

Different models have been considered to capture the backhaul rate constraints~\cite{SSS08PIMRC}. A \emph{central} coordination model assumes that all BSs are connected to a central station through finite-capacity backhaul links~\cite{SSP09TITAug,HKD11TSP}. The model is simple and thus amenable to theoretical analysis, but may have scalability and robustness issues in practical implementation.  A \emph{distributed} backhaul model assumes that the BSs can communicate with one another via finite-capacity backhaul links, which may be confined to a neighborhood~\cite{MaF07,MaF11TWC}.
The backhaul capacity may be shared among the BSs for successive interference cancellation to achieve rates within a constant number of bits to the theoretical limits~\cite{zhou2013uplink}. A third approach is to consider \emph{clusters} of BSs that cooperate tightly, possibly with coordination across the clusters as well to a lesser degree~\cite{Ven07,PGH08,ZCA09}.  

\subsection{Proposed Approach}
Our approach is to design a linear combiner for joint decoding with the dual objectives of minimizing the mean-square error (MSE) of the combiner output, as well as the required backhaul rate. To perform joint decoding, a BS collects received signal samples at the neighboring BSs and combine them using a linear equalizer. The amount of the backhaul traffic\footnote{ In this work, the control signaling overhead is not explicitly captured in the backhaul overhead calculation, since it is negligible in comparison to the overhead due to sharing analog signal samples for joint decoding. The actual backhaul traffic rate, say in Mbps, depends on many implementation details such as the sampling frequency, quantization resolution, as well as various types of overhead. Therefore, in this work, what we term as the backhaul traffic captures the number of packets€ of samples transported from one BS to another BS through the backhaul network per unit time. Depending on how many samples (e.g. samples from multiple antennas at each BS) quantized with how many bits per sample are contained in one packet, the actual number in terms of Mbps will change. Also, due to this unspecified scaling, the backhaul traffic discussed in this work will not be represented in terms of particular units.} incurred in this process is proportional to the number of the neighboring BSs from which the samples are collected. The number of cooperating BSs, in turn, is the same as the number of non-zero entries in the equalizer coefficient vector. To determine which coefficients to allow to be non-zero under the minimization objective requires a combinatorial search in principle. Our pragmatic approach is to adopt a convex relaxation of the problem in the spirit of the recent compressive sensing literature in order to design {\em sparse} equalizers~\cite{Tib96}.

Both distributed and clustered cooperation models are considered. In the distributed cooperation setup, each BS communicates directly with the cooperating BSs to obtain the signal samples. In the clustered scenarios, the BSs are partitioned into a number of non-overlapping clusters, and the signal samples are fully shared among the BSs in a cluster for joint decoding of the served MSs~\cite{PGH08,ZCA09}. It is assumed that the inter-cluster backhaul traffic constitutes the bottleneck, rather than the intra-cluster backhaul traffic. When the clusters are given and fixed, the equalizer weights corresponding to the inter-cluster feedback are sparsified. To solve jointly for dynamic clusters and equalizer weights, a spectral clustering approach is derived~\cite{Lux07}.

\subsection{Related Work}
MCP in the downlink with limited backhaul rates has been investigated in the literature. Cooperative downlink transmission with per-BS power and Quality-of-Service (QoS) constraints was studied in~\cite{ZQL13}. A clustered cooperation scheme with linear processing in the downlink was proposed in~\cite{ZCA09}. In the context of multi-cell MIMO heterogeneous networks, joint BS clustering and beamformer design for downlink transmission were considered in~\cite{Luo2013downlink_beamforming, liao2013base}. An energy-constrained beamformer design and BS-MS association for uplink and downlink were considered in \cite{luo2014downlink}. Distributed precoder design and BS selection in a game theoretic framework were proposed in~\cite{Luo2013uplink_precoder}. A semidefinite relaxation-based approach for backhual-limited cooperation in the downlink was proposed in~\cite{ZhV14}. Particle swarm optimization was used for zero-forcing-type beamformer design in~\cite{lakshmana2012partial}. Our work focuses on the uplink and does not require coordination with MSs, with all MCP burden placed on the BSs.

Direct exchanges of the signal samples between the BSs were considered in the context of 3GPP LTE systems in~\cite{HFG09}, where distributed cooperation without central control was advocated. Overlapping clusters of BSs were elected based on proximity for uplink MCP in~\cite{Ven07}. A greedy algorithm for dynamic clustering was proposed to maximize the uplink sum-rate in~\cite{PGH08}. Successive interference cancellation was adopted under limited backhaul traffic in~\cite{zhou2013uplink}, and cooperative group decoding was considered in a similar setting in~\cite{LWZ14}. Here, our intention is to concentrate on simple linear processing but address the backhaul traffic volume issue in both distributed and clustered cooperation settings in a consistent framework, and also derive distributed algorithms for scalable implementation.

Compared to our conference precursor~\cite{kim2012backhaul}, decentralized implementation of the proposed distributed and clustered cooperation schemes is developed in the present work. Decentralized computation of spatial equalizers and clusters makes MCP scalable to large networks, robust to isolated points of failure, and more resource-efficient than centralized implementation, since pieces of the overall problem are solved concurrently at different BSs, coordinated by peer message exchanges. Decentralized implementation of the component algorithms, such as group-sparse regression, eigenvector computation for network Laplacians, and k-means clustering, is also discussed. In addition, extensive simulations are performed to verify the performance, including the use of multiple antennas, user fairness, and dynamic clustered cooperation scenarios.

The rest of the paper is organized as follows. Sec.~\ref{sec:model} introduces the system model. Sec.~\ref{sec:distr} develops backhaul-constrained distributed cooperation, while  Sec.~\ref{sec:cluster} presents static and dynamic clustered cooperation. Decentralized implementation is the subject of Sec.~\ref{sec:decentr}. Numerical tests are performed in Sec.~\ref{sec:test}, and conclusions are provided in Sec.~\ref{sec:conc}.

\section{System Model}
\label{sec:model}
\subsection{Uplink Signal Model}
\label{ssec:signal}
Consider a multi-cell network uplink with $N_B$ BSs and $N_U$ MSs (users). Each BS $b \in \ccalB := \{1,2,\ldots, N_B\}$ is assumed to be equipped with $A$ antennas. Thus, the total number of BS antennas in the network is $N_A := A N_B$. The set of antennas that belong to BS $b$ is denoted as $\ccalA_b$. Each MS user $u \in \ccalU := \{1,2,\ldots,N_U\}$ possesses a single antenna. The serving BS of the $u$-th MS is denoted as $b(u)$. The set of MSs served by BS $b$ is denoted as $\ccalU_b := \{u \in \ccalU : b(u) = b\}$.

Let $x_u$ denote the symbol transmitted by the $u$-th MS in a particular time slot and (sub-)band. Define $\bbx := [x_1,x_2,$ $\ldots,x_{N_U}]^T$, where $^T$ stands for transposition. Similarly, the signal samples received by the $N_A$ BS antennas are represented as $\bby := [y_1,y_2,\ldots,y_{N_A}]^T$, where $y_a$ for $a \in \ccalA := \{1,2,\ldots,N_A\}$ denotes the sample taken by the $a$-th antenna. The flat-fading channel matrix $\bbH \in \mathbb{C}^{N_A \times N_U}$ has its $(a,u)$-entry equal to the complex channel coefficient from the $u$-th MS to the $a$-th BS antenna. Thus, the input-output relationship for the uplink can be expressed compactly as
\begin{align}
\bby = \bbH \bbx + \bbn
\end{align}
where $\bbn$ represents additive noise and possible interference from outside the network. Vectors $\bbx$ and $\bbn$ are assumed to be zero-mean complex random vectors with covariance $\bbI$, which can be justified through prewhitening and normalization, as well as by absorbing MS transmit-powers into $\bbH$. It is also assumed that $\bbx$ and $\bbn$ are uncorrelated.

With MCP, multiple BSs cooperate to decode $\bbx$. For simplicity of implementation, linear receiver processing is considered. If all $N_A$ BS antennas in the network could fully cooperate, an estimate $\hat \bbx$ of $\bbx$ could be obtained using a linear equalizer $\bbW \in \mathbb{C}^{N_U \times N_A}$ as
\begin{align}
\hat \bbx = \bbW \bby
\end{align}
where $\hat \bbx := [\hat x_1,\ldots,\hat x_{N_U}]^T$. Thus, in order to obtain the estimate $\hat x_u$ of $x_u$ of MS $u \in \ccalU$, BS $b(u)$ would need to collect samples received by other BS antennas, and compute $\bbw_u^T \bby$, where $\bbw_u^T$ denotes the $u$-th row of $\bbW$. A widely used linear equalizer, also adopted here, is the linear minimum mean-square error (LMMSE) one, given by
\begin{align}\label{eq:W_LMMSE}
\bbW_\mathrm{lmmse} &= 
(\bbH^\ccalH \bbH + \bbI)^{-1} \bbH^\ccalH
\end{align}
where $^\ccalH$ denotes Hermitian transposition. 

\begin{figure}
\centering
\begin{tabular}{cc}
\includegraphics[scale=0.7]{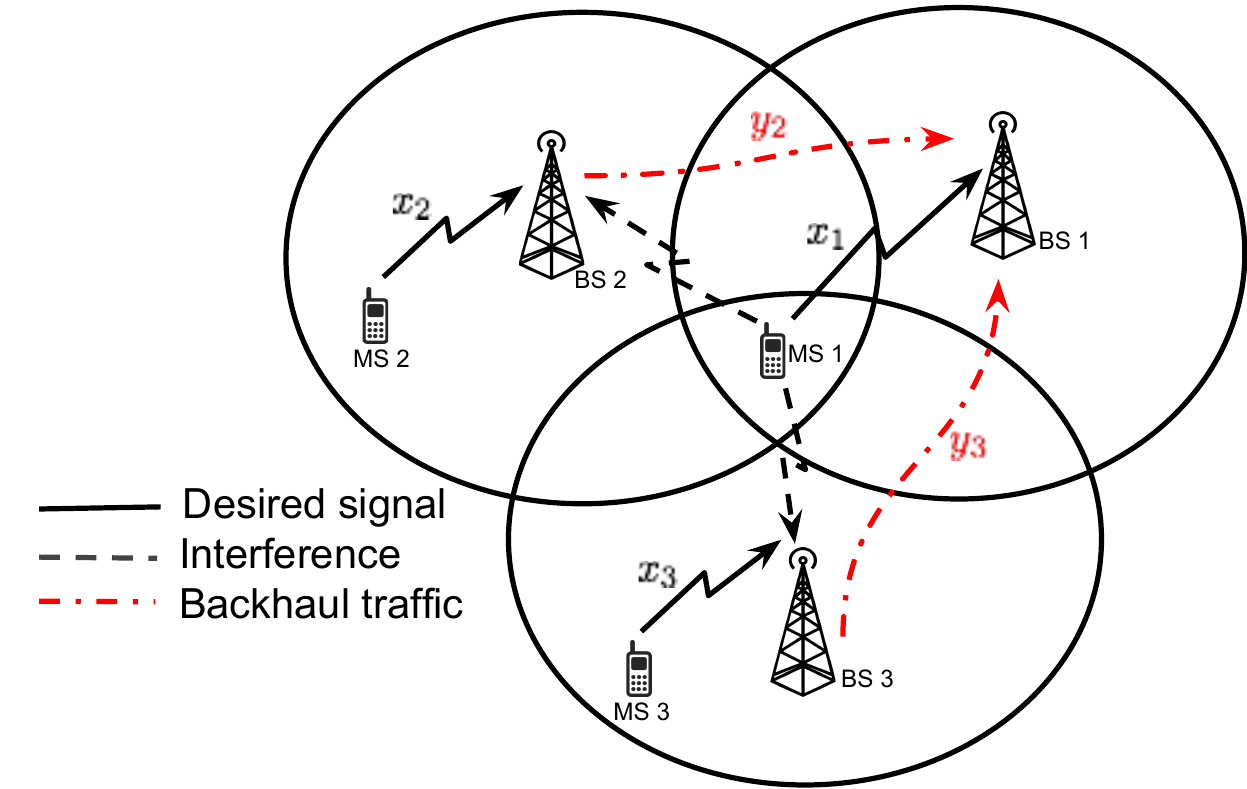}&
\includegraphics[scale=0.5]{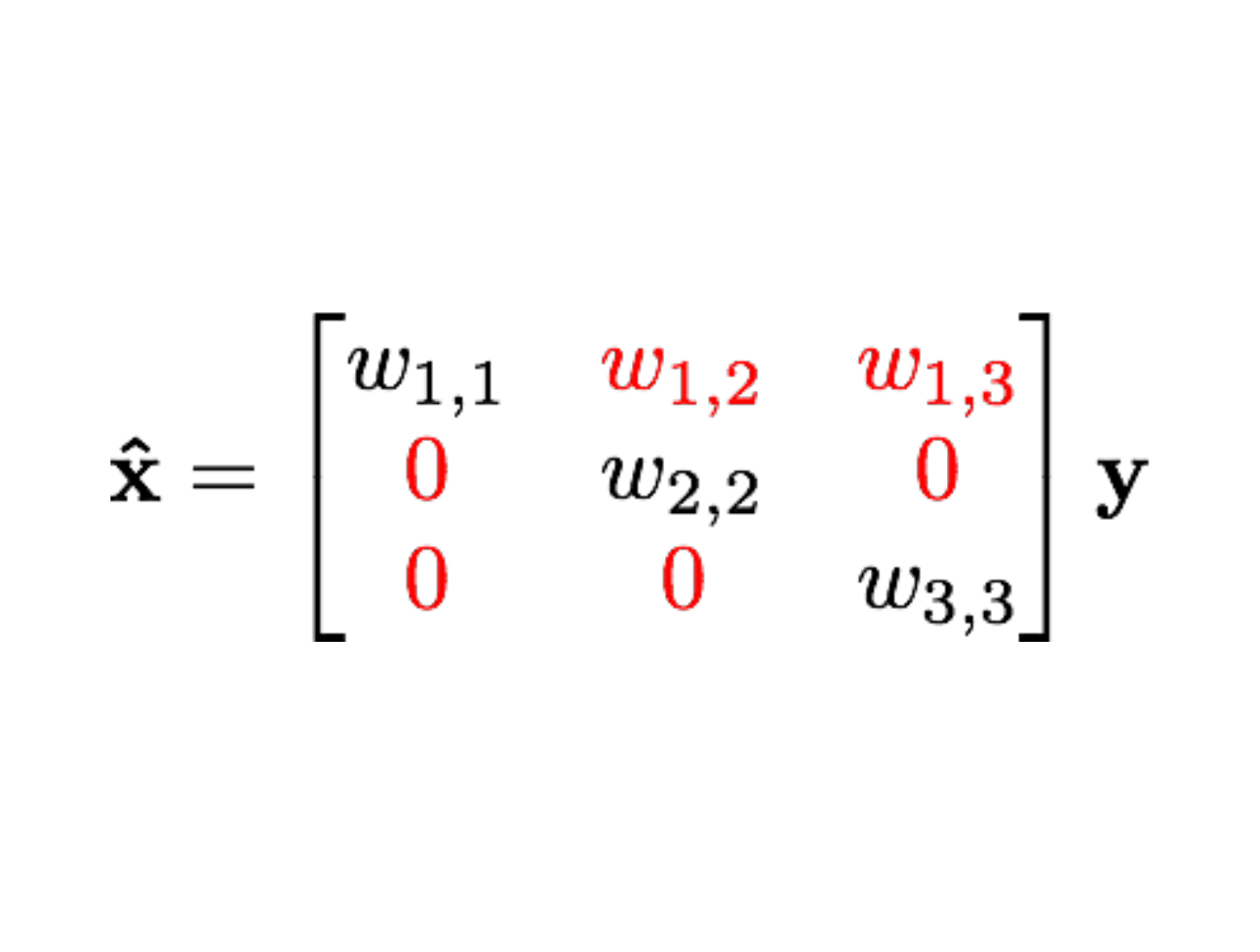}
\end{tabular}
\caption{An example MCP model with 3 single-antenna BSs. MS~1's transmission causes interference at BSs~2 and~3, but these BSs can cooperate by backhauling their received samples to BS~1, where joint decoding of MS~1 is performed. The backhaul traffic manifests itself as the non-zero off-diagonal entries in the first row of $\bbW$.}
\label{fig:ex_model}
\end{figure}

\subsection{Distributed Cooperation Model}
\label{ssec:distr-coop}
In the simplest cooperation model, henceforth termed distributed cooperation, each BS collects from cooperating BSs the received samples necessary for decoding its served MSs~\cite{HFG09}. The set of cooperating BSs may overlap, in the sense that the set of BSs helping BS $b$ may have nonempty intersection with the set of BSs helping BS $b'$ for $b \neq b'$~\cite{Ven07,NgH10}. The BSs share the samples over the backhaul network. If unlimited backhaul traffic were allowed, every BS could collect samples from all other BSs to perform full cooperation. Since the backhaul capacity is limited in practice, each BS must judiciously choose the set of BSs whose samples are most helpful in decoding its intended MSs.

Given the equalizer matrix $\bbW$, the amount of backhaul traffic can be characterized. To get the basic idea, consider again the case of decoding MS $u$'s signal by forming $\bbw_u^T \bby$. Clearly, collecting the entries in $\bby$ that are measured by BSs other than the serving BS $b(u)$ must be conveyed to BS $b(u)$ through the backhaul links. These are the samples $\{y_a\}$ for $a \in \ccalA \backslash \ccalA_b$. However, if the $a$-th entry $w_{ua}$ of $\bbw_u^T$ is zero, then the corresponding entry $y_a$ need not be collected. Therefore, the actual backhaul traffic amount incurred is proportional to the number of nonzero entries in $\{w_{ua}\}_{a \in \ccalA \backslash \ccalA_{b(u)}}$.

The idea can be applied straightforwardly in the case of single-antenna BSs; that is when $A=1$. Assume that $N_B = N_U = N_A$ with $b(u) = u$ for all $u \in \ccalU$, and that the $b$-th antenna belongs to BS $b$; that is $\ccalA_b = \{b\}$ for $b \in \ccalB$. Then, under the distributed cooperation model, the amount of backhaul traffic is proportional to the number of nonzero off-diagonal entries of $\bbW \in \mathbb{C}^{N_B \times N_B}$.  An illustration for the case with 3 BSs is provided in Fig.~\ref{fig:ex_model}.

In the case of multi-antenna BSs with $A > 1$, each BS may well serve more than one MS. There are two considerations to take into account when estimating the amount of backhaul traffic in the case of multi-antenna cell sites. 
\begin{enumerate}
	\item[c1)] Once BS $b$ has collected samples from a cooperating BS~$b'$ to decode a user $u \in \ccalU_b$, no additional backhaul traffic is incurred even if these samples are used for decoding other MSs $u' \neq u$ that are in $\ccalU_b$. In other words, once the $(u,a')$-entry of $\bW$ is determined to be nonzero, where $u \in \ccalU_b $ and $a' \notin \ccalA_b$, one can improve the decoding performance without incurring additional backhaul traffic by allowing $(u',a')$-entries of $\bW$ to be nonzero for $u' \in \ccalU_b$.
	
	\item[c2)] When a signal sample received from antenna $a \in \ccalA_b$ is collected, it may be desirable to collect together the samples from all the antennas in $\ccalA_b$ associated with that BS in order to economize the overhead due to backhaul packet headers and to maximize the MCP benefit.\footnote{As all-IP networking is advocated for cellular backhauling, the backhaul traffic is often routed using packet switching, which incurs header overhead. Assuming that the multiple antennas at a BS receive the MS signal at similar strengths, it is reasonable to collect all the samples from the multiple antennas together to maximize the MCP benefit. In this case, although the payload size increases linearly in the number of antennas, the header overhead is constant, as the samples may be aggregated into a single packet.}
\end{enumerate}

A way to succinctly capture c1) and c2) is to consider the set
\begin{align}
\left\{ \sum_{u \in \ccalU_b} \sum_{a \in \ccalA_{b'}} |w_{ua}|^2 \Bigg| b,b' \in \ccalB,\ b \neq b' \right\} \label{eq:tildeW_set}
\end{align}
where it is worth noting that in order for an element to be zero, all squared terms in the sum must be zero. Thus, the backhaul traffic incurred is proportional to the number of nonzero elements in the set in \eqref{eq:tildeW_set}.

\subsection{Clustered Cooperation Model}
In the clustered cooperation, BSs are partitioned into a number of non-overlapping clusters, and the received samples are fully shared among the BSs belonging to a cluster so that the MSs are served jointly~\cite{PGH08,ZCA09}. Since the clusters do not overlap, the cluster-edge users may suffer from interference~\cite{CRP08}. In some cases, natural clustering could emerge by the deployment constraints of the network, such as the hierarchical routing architecture of the backbone network, or the geographical proximity of the cell sites. In other cases, dynamic formation and adaptation of clusters may be beneficial. In the latter case, clusters themselves can be optimized based on channel gains or traffic patterns~\cite{PGH08}. 

It is assumed that inter-cluster backhaul traffic is more costly to the backhaul network than intra-cluster traffic. For example, when multiple cells are controlled by a single cell tower, backhauling between these cell sites are virtually free. Similarly, the particular routing topology of the backhaul implementation may dictate certain backhauling paths are much cheaper than others. Inter-cluster backhaul traffic may pass through multiple layers of routing hierarchy, incurring higher signaling overhead. The precise mathematical backhaul traffic characterization will be presented in Sec.~\ref{sec:cluster}. Next, the two modes of MCP operation will be studied one by one.

\section{Distributed Cooperation}
\label{sec:distr}

\subsection{Sparse LMMSE Equalizer}
\label{ssec:sLMMSE}
For simplicity of exposition, consider first the case with $A = 1$, $N_B = N_U = N_A$, $\ccalU_b = \{b\}$ and $\ccalA_b = \{b\}$ for all $b \in \ccalB$, as in Sec.~\ref{ssec:distr-coop}. Since the amount of backhaul traffic is proportional to the number of nonzero off-diagonal entries of $\bW$, a natural approach to obtain a backhaul-constrained MCP solution is to promote sparsity in the off-diagonal entries of $\bW$~(that is, favor many zero entries in the off-diagonal positions with minimal sacrifice in the MCP gain). Inspired by recent advances in compressive sensing techniques, the idea here is to optimize the cost functions related to the MMSE criterion, augmented with sparsity-promoting regularization terms~\cite{Tib96,CDS98}.

Let $\tilde \bW \in \mathbb{R}^{N_B \times N_B}$ denote the matrix with zero diagonal entries, and the off-diagonal entries equal to the absolute values of the corresponding entries in $\bbW$.%
\footnote{Since the entries of $\bbW$ are complex numbers, setting an entry to zero amounts to setting both the real and the imaginary parts of the entry to zero. Considering the absolute value of the complex entry satisfies this requirement, and leads later to the group sparsity formulation quite naturally~\cite{YuL06}. More discussion on group sparsity will be provided in Sec.~\ref{ssec:multiant}.}
The LMMSE equalizer seeks to minimize the mean-square error (MSE) $\E \{||\hat \bx - \bx||_2^2\} = ||\bI - \bW \bH||_F^2 + ||\bW ||_F^2$. Thus, one can penalize large values of $\ell_0$-``norm" of $\tilde \bW$ as in
\begin{align}
\min_{\bbW} \|\bbI - \bbW \bbH \|_F^2 + \|\bbW \|_F^2 + \lambda \|\tilde \bbW\|_0 \label{eq:MMSE-l0}
\end{align}
where $\|\tilde \bbW\|_0$ denotes the total number of nonzero entries in $\tilde \bbW$, $\|\cdot\|_F$ the Frobenius norm, and $\lambda$ controls the amount of backhaul traffic incurred by the solution.

Since the criterion in~\eqref{eq:MMSE-l0} is nonconvex, to obtain a solution efficiently, we advocate replacing the $\ell_0$-``norm" with the $\ell_1$-norm, which offers the tightest convex relaxation of the $\ell_0$ counterpart. Thus, the relevant optimization problem is
\begin{align}
\min_{\bbW} ||\bbI - \bbW \bbH||_F^2 + ||\bbW ||_F^2 + \lambda ||\tilde \bbW||_1
\end{align}
which is equivalent to
\begin{align}
\min_{\bbW} \left\| \bar \bbI - \bbW \bar \bbH \right\|_F^2 + \lambda ||\tilde \bbW||_1 \label{eq:MMSEprob}
\end{align}
where $\bar \bbI := \big[ \bI \ \ \mathbf{0}  \big]$, $\bar \bbH := \big[ \bH \ \ \bI \big]$, and $\| \tilde \bbW \|_1$ is the sum of the (absolute) values of all entries of $\tilde \bbW$.

\begin{remark}
In practice, one must choose an appropriate value of $\lambda$ to obtain the desired backhaul traffic amount. For a large enough value of $\lambda$, the entire matrix $\bbW$ will be equal to zero. Thus, a viable strategy is to first compute an upper bound $\lambmax$ for the $\lambda$ value that yields the all-zero solution, and to search in the interval $[0,\lambmax]$. $\lambmax$ can be computed in a closed form. The least angle regression (LARS) algorithm can be used to efficiently calculate the entire regularization path, which is the mapping between $\lambda$ and the corresponding sparsity level (backhaul traffic amount)~\cite{efron2004least}.
\end{remark}

\begin{remark} In our formulation, the total backhaul traffic amount is considered, and the backhaul traffic per BS is not individually controlled. However, the latter can be attained in the same framework by introducing different weights for individual rows of $\bbW$. For example, employing weights $\{\lambda_b\}$, one can consider
\begin{align}
\min_{\{\bbw_b\}_{b=1}^{N_B}} \sum_{b=1}^{N_B}\left[\|\bar \bbI(b,:) - \bbw_b^T \bH\|_F^2 + \lambda_b \|\tilde \bbw_b\|_1 \right]
\end{align}
where $\bar \bbI(b,:)$ represents the $b$-th row of $\bar \bbI$, $\tilde \bbw_b^T$ the $b$-th row of $\tilde \bbW$, and $\lambda_b$ captures the severity of the backhaul constraint for BS $b$.
\end{remark}

\subsection{Multi-Antenna Case Using Group Sparsity}
\label{ssec:multiant}
Considerations c1) and c2) discussed in Sec.~\ref{ssec:distr-coop} for characterizing the backhaul traffic naturally impose certain structural constraints in sparsifying $\bbW$ in the multi-antenna case. That is, for any pair of BSs $b$ and $b'$ with $b \neq b'$, the weights $w_{ua}$ for $u \in \ccalU_b$ and $a \in \ccalA_{b'}$ should be constrained to be either all zero or all nonzero.

Such a structure can be effected using penalty terms promoting {\em group sparsity}~\cite{YuL06}. When a vector $\bbw \in \mathbb{R}^N$ is partitioned to $J$ subvectors as $\bbw := [\bbw_1^T,\bbw_2^T,\ldots,\bbw_J^T]^T$ with $\bbw_j \in \mathbb{R}^{N_j}$ and $\sum_{j=1}^J N_j = N$, rather than pursuing the sparsity of the individual entries in $\bbw$ independently by using a penalty term of the form $\lambda \|\bbw\|_1$, one can encode group structures by using $\lambda \sum_{j=1}^J \|\bbw_j\|_2$. This gives rise to $\bbw_j$'s that are either entirely zero or nonzero.

Under these considerations, one can replace the penalty term in~\eqref{eq:MMSEprob} with
\begin{align}
\lambda \sum_{b,b' \in \ccalB: b \neq b'} \sqrt{\sum_{u \in \ccalU_b} \sum_{a \in \ccalA_{b'}} |w_{ua}|^2 } \label{eq:penalty-distr}
\end{align}
where $w_{ua}$ is the $(u,a)$-entry of $\bW$. The first sum inside the square root collects the entries for all served MSs in BS $b$, and the second one for all the antennas in a cooperating BS $b'$. Another way of expressing this is by defining $\tilde \bbW \in \mathbb{R}^{N_B \times N_B}$ with $(b,b')$-entry $\tilde w_{bb'}$ for $b,b' \in \ccalB$ given as
\begin{align}
\tilde w_{bb'} = \begin{cases} \sqrt{\sum_{u \in \ccalU_b} \sum_{a \in \ccalA_{b'}} |w_{ua}|^2 } & \textrm{if } b \neq b' \\
0 & \textrm{otherwise}. \end{cases} \label{eq:def-tildeW}
\end{align}
Then, $\|\tilde \bbW\|_0$ represents the amount of backhaul traffic, and the relaxed problem \eqref{eq:MMSEprob} with this new $\tilde \bbW$ can be solved to obtain a group-sparse LMMSE equalizer efficiently. Note that \eqref{eq:def-tildeW} coincides with the definition of $\tilde \bbW$ in Sec.~\ref{ssec:sLMMSE} when $A = 1$, $N_B = N_U$, $\ccalU_b = \{b\}$, and $\ccalA_b = \{b\}$ $\forall b \in \ccalB$.

\section{Clustered Cooperation}
\label{sec:cluster}

\subsection{Static Clustering}
\label{ssec:fixed}
In the clustered cooperation scenario, the set of BSs is partitioned into $N_C$ non-overlapping clusters $\{\cV_c\}_{c=1}^{N_C}$, i.e.,
\begin{align}
\bigcup_{c=1}^{N_C} \cV_c = \cB \quad \textrm{ and } \quad \cV_c \cap \cV_{c'} = \varnothing \textrm{ for } c \neq c'. \label{eq:cl-constr}
\end{align}

The BSs belonging to the same cluster fully cooperate by sharing the received signal samples. This can be done in a distributed fashion via broadcasting the samples within each cluster through the backhaul network, or by having a central unit per cluster such as the cluster head, which collects the samples from the cluster members for further processing. In the latter case, with a BS in each cluster elected as the cluster head, the intra-cluster traffic of cluster $\cV_c$ can be modeled as proportional to $|\cV_c|-1$, where~$|\cdot|$ denotes the set cardinality. In the former case, the intra-cluster backhaul traffic is proportional to $|\cV_c|(|\cV_c|-1)$, since each BS in a cluster needs samples from all the other BSs in the same cluster. Note that electing cluster heads may incur additional control overhead and implementation issues.

In {\em static} clustered cooperation, the clusters $\{\cV_c\}_{c=1}^{N_C}$ are given and fixed, regardless of the system operating conditions such as the channel realizations. Static clustering is appealing thanks to its simplicity of implementation. Clustering based on the geographical proximity of cells or the network architectural constraints is natural in this scenario. Cells that are within a fixed radius were clustered together in~\cite{Ven07}. The distributed antenna system (DAS) connected to a central control unit makes a straightforward cooperating cluster~\cite{PDF08}.

Unlike the distributed cooperation discussed in Sec.~\ref{sec:distr}, clustered cooperation can have an edge effect. That is, while the users located in the center of a cluster can clearly benefit from the MCP, the users at the cluster edges can suffer from excessive interference~\cite{CRP08}. Judicious inter-cluster cooperation can help mitigate such a drawback. To this end, it is necessary to assess the inter-cluster traffic overhead.

The inter-cluster backhaul traffic is proportional to the number of nonzero $(b,b')$-entries of $\tilde \bbW$, when BSs $b$ and $b'$ belong to different clusters, namely, $b \in \cV_c$ and $b' \in \cV_{c'}$ with $c \neq  c'$. By relaxing this cardinality-based characterization using the $\ell_1$-norm, a suitable regularizer is
\begin{equation}
\lambda \sum_{(c,c'): c \neq c'} s(\cV_c,\cV_{c'}) \label{eq:penalty-cl}
\end{equation}
where
$s(\cV_c,\cV_{c'}) := \sum_{b \in \cV_c} \sum_{b' \in \cV_{c'}} \tilde w_{bb'}$.
Note that in the special case of singleton clusters, i.e., if $N_C = N_B$ and $\ccalB_b = \{b\}$ for all $b \in \ccalB$, then \eqref{eq:penalty-cl} reduces to \eqref{eq:penalty-distr}.

It is useful to view the backhaul penalty in~\eqref{eq:penalty-cl} from the perspective of graph cuts.\footnote{A graph is an ordered pair of a set of vertices and a set of edges. A weighted graph is a graph, whose edges are assigned with real numbers (weights). A cut of a weighted graph is the sum of the weights associated with the edges, whose endpoints lie in different partitions of vertices.} Consider a directed graph with the set of vertices comprising the BSs $\ccalB$, and the set of directed edges $\ccalE := \{(b,b'):b,b'\in \ccalB\}$ connecting the vertices. The edge weight associated with edge $(b,b')$ is given by $\tilde w_{bb'}$ for $b,b' \in \ccalB$; i.e., $\tilde \bbW$ is the affinity matrix. Then, $s(\ccalB_c, \ccalB_{c'})$ is the sum of edge weights that connect the vertices in $\ccalB_c$ to the vertices in $\ccalB_{c'}$. Thus, \eqref{eq:penalty-cl} is proportional to the graph cut on $\ccalG := (\ccalB, \ccalE, \tilde \bbW)$ induced by the set of clusters $\{\ccalB_c\}$, defined as
\begin{align}
\cut(\{\ccalB_c\}) := \sum_{c=1}^{N_C} s(\ccalB_c, \overline \ccalB_c) = \sum_{(c,c'):c \neq c'} s(\ccalB_c,\ccalB_{c'})
\end{align}
where $\overline \ccalB_c := \cup_{c' \neq c} \ccalB_{c'}$. Thus, static clustered MCP solves
\begin{align}
	\min_{\bbW} ||\bar \bbI - \bW \bar \bbH||_F^2 +  \lambda \cdot \cut(\{\ccalB_c\}) \label{eq:statcl} 
\end{align}
which is a convex optimization problem. Note that in the special case of singleton clusters, i.e., if $N_C = N_B$ and $\ccalB_b = \{b\}$ for all $b \in \ccalB$, then \eqref{eq:statcl} reduces to the distributed cooperation problem in~\eqref{eq:MMSEprob}.

\subsection{Dynamic Partitioning Using Spectral Clustering}
\label{ssec:spec-cl}
When the clusters are {\em not} given a priori, or when one is allowed to adapt the clusters dynamically, it is of interest to determine the partitioning that provides the best tradeoff between the system performance and the feedback overhead, possibly with limited inter-cluster coordination. However, finding the best $\{\cV_c\}$ using the regularizer \eqref{eq:penalty-cl} leads to clusters of severely unbalanced sizes. For instance, taking $\cV_1 = \cB$ and $\cV_c = \varnothing$ for $c=2,\ldots,N_C$ is the degenerate optimum for~\eqref{eq:penalty-cl} with full cooperation and zero inter-cluster traffic. Obviously, such unbalanced cluster sizes are undesirable as the intra-cluster traffic can become excessive.

To obtain clusters with balanced sizes, one may consider the {\em ratio cut} cost used in spectral graph partitioning~\cite{HaK92}:
\begin{align}
\rcut(\{\ccalB_c\}) := \sum_{c=1}^{N_C} \frac{s(\cV_c, \overline{\cV}_c)}{|\cV_c|} \;.\label{eq:rcutdef}
\end{align}
Here, the clusters of small sizes are penalized as $|\cV_c|$ is in the denominator. Thus, a joint sparse LMMSE equalization and clustering problem is formulated as
\begin{align}
\min_{\bW, \{\cV_c\}_{c=1}^{N_C}} \|\bar \bI - \bW \bar \bH\|_F^2 + \lambda \cdot \rcut(\{\ccalB_c\}) \label{eq:prob-rcut}
\end{align}
where $\{\ccalB_c\}$ must satisfy \eqref{eq:cl-constr}. 

Even with fixed $\bbW$, problem \eqref{eq:prob-rcut} is known to be NP-complete~\cite{HaK92}. To obtain an approximate solution, it is first noted that with fixed $\bbW$, optimization w.r.t. $\{\ccalB_c\}$ is precisely the minimum cut problem on graph $\ccalG$, whose relaxation is what spectral clustering algorithms aim to solve~\cite{Lux07,MeP07}. On the other hand, given $\{\ccalB_c\}$, optimization w.r.t. $\bbW$ is a convex problem, as noted in Sec.~\ref{ssec:fixed}. Thus, an iterative approach is pursued, where one alternately optimizes over $\bW$ and $\{\ccalB_c\}$ until no more reduction in the cost is obtained.

Spectral clustering has been mostly applied to undirected graphs~\cite{Lux07}. To perform spectral clustering on directed graphs, some care must be taken~\cite{MeP07}. Define matrix $\bbD \in \mathbb{R}^{N_B \times N_B}$ to be a diagonal matrix with $b$-th diagonal entry $\sum_{b' \in \ccalB} \tilde w_{bb'}$. Define also $\bbL := \bbD - \tilde \bbW$. Consider an $N_B \times N_C$ cluster indicator matrix $\bbPhi$ whose $(b,c)$-entry is defined as
\begin{align}
\phi_{bc} = \begin{cases} \frac{1}{\sqrt{|\ccalB_c|}} & \textrm{ if } b \in \ccalB_c\\
0 & \textrm{ otherwise}\end{cases} \;. \label{eq:phi-constr}
\end{align}
As clusters are non-overlapping, it can be seen that $\bbPhi^T  \bbPhi = \bbI$. Then, it follows that [cf. \eqref{eq:rcutdef}]
\begin{align}
\rcut(\{\ccalB_c\}) = \tr(\bbPhi^T \bbL \bbPhi)\;.
\end{align}
Thus, a minimum cut graph clustering problem based on the ratio cut objective can be stated as
\begin{align}
\min_{\bbPhi} \tr (\bbPhi^T \bbL \bbPhi) \textrm{ subject to } \bbPhi^T \bbPhi = \bbI \textrm{ and \eqref{eq:phi-constr}}. \label{eq:sp-cl}
\end{align}
Spectral clustering drops constraint~\eqref{eq:phi-constr} such that $\bbPhi$ takes real-valued (rather than binary) entries. 
However, since $\bbL$ is not symmetric, one cannot apply the Rayleigh-Ritz theorem to the resulting relaxed problem directly. Noting that $\tr(\bbA) = \tr(\bbA^T)$,  one can obtain a formulation equivalent to~\eqref{eq:sp-cl} (except for the dropped constraint~\eqref{eq:phi-constr}) as
\begin{align}
\min_{\bbPhi} \tr (\bbPhi^T \tilde \bbL \bbPhi) \textrm{ subject to } \bbPhi^T \bbPhi = \bbI \label{eq:sp-cl-2}
\end{align}
where $\tilde \bbL := (\bbL + \bbL^T)/2$ is symmetric. The optimal solution to \eqref{eq:sp-cl-2} is given by setting the columns of $\bbPhi$ as the eigenvectors of $\tilde \bbL$ corresponding to the $N_C$ smallest eigenvalues. Clusters $\{\ccalB_c\}$ can then be found by running the k-means algorithm on the rows of $\bbPhi$. The overall algorithm is listed in Table~\ref{tab:alg-dynamic}. 

\begin{table}[t]
\centering
\begin{tabular}{|l|}
\hline
Input: $\lambda$, number of clusters $N_C$, initial equalizer $\bbW_0$, small $\varepsilon \ge 0$\\
Output: $\{\ccalB_c\}_{c=1}^{N_C}$, $\bbW$\\
1: Initialize $\bbW \leftarrow \bbW_0$\\
2: Repeat:\\
3: \qquad Compute $\tilde \bbW$ per~\eqref{eq:def-tildeW}.\\
4: \qquad Set $\bbD \leftarrow \textrm{diag}(\tilde \bbW {\bf 1})$ (${\bf 1}$ is the all-one vector)\\
5: \qquad Set $\bbL \leftarrow \bbD - \tilde \bbW$ and $\tilde \bbL \leftarrow (\bbL + \bbL^T) / 2$\\
6: \qquad Let $\bbphi_c$ be the eigenvector of $\tilde \bbL$ corresponding to\\
\qquad\qquad the $c$-th smallest eigenvalue for $c=1,\ldots,N_C$.\\
7: \qquad Run the k-means algorithm on the rows of \\
\qquad\qquad $\bbPhi := [\bbphi_1,\ldots,\bbphi_{N_C}]$ to obtain clusters $\{\check \ccalB_c\}$.\\
8: \qquad $\check \bbW = \arg \min_{\bW} \|\bar \bI - \bW \bar \bH\|_F^2 + \lambda \cdot \rcut(\{\check \ccalB_c\})$\\
9: \qquad If the decrease in the objective in line 8 is less than $\varepsilon$, stop.\\
10: \qquad \hspace{-8pt} Set $\bbW \leftarrow \check \bbW$ and $\{\ccalB_c\} \leftarrow \{\check \ccalB_c\}$.\\
\hline
\end{tabular}
\caption{Joint MCP and dynamic partitioning.}
\label{tab:alg-dynamic} \vspace{-10mm}
\end{table}

\section{Decentralized Implementation}
\label{sec:decentr}
\label{sec:dist_stat}
The algorithms developed so far require knowledge of the full channel matrix $\bbH$. In practice, a particular BS might not have access to the full $\bbH$, but only to the local channel gains, such as the gains between itself and the nearby MSs including the ones served by the BS. This essentially provides one or part of a few rows of $\bbH$. A centralized implementation would need to collect all the rows of $\bbH$ from individual BSs at a central processor, solve the appropriate MCP problem, and feed back the solution to the relevant BSs. However, such an implementation may not be very scalable, but rather vulnerable as the failure of the central processor can affect the entire system. Furthermore, the feedback overhead can be significant. A naive decentralized implementation would simply broadcast $\bbH(b,:)$ across all BSs, whereupon each BS can independently solve the relevant MCP problems for the entire system. Clearly, however, the computational resources are wasted. Moreover, in the case of dynamic clustering, the solution to the spectral clustering may not be unique when $\tilde \bbL$ has repeated eigenvalues, and the results of the k-means algorithm may be different from BS to BS.

In view of these considerations, the following features are desired for the decentralized implementation.
\begin{itemize}
\item Only locally available channel state information should be used.

\item The computation should be efficiently done by having individual BSs solve only the subproblems of the entire problem.

\item There must be a consensus on the solution across the BSs at the end.
\end{itemize}
Next, decentralized algorithms for static and dynamic versions of MCP are derived, satisfying the aforementioned requirements.\footnote{The decentralized algorithm for static clustered MCP covers the case of distributed cooperation as well. This is because distributed MCP is a special case of static clustered MCP, where each cluster contains a single BS and the number of clusters is equal to number of BSs.}

\subsection{Decentralized Algorithm for Statically Clustered MCP}
In this section, a decentralized algorithm for solving \eqref{eq:statcl} is developed. First, consider a communication graph $\ccalG_c := (\ccalB, \ccalE_c)$, where the BSs $b$ and $b'$ can communicate whenever an undirected edge $(b,b') \in \ccalE_c$. Not to count the same edge twice, assume without loss of generality that $b < b'$ for all $(b,b') \in \ccalE_c$. 
It is assumed that $\ccalG_c$ is connected, which means that there is always a path from any node to any other node. To simplify the exposition, single-antenna BSs ($A = 1$) are assumed in the sequel. However, it is straightforward to extend the algorithm to the case with $A > 1$. Finally, it is postulated that each BS knows the column of $\bbH$ corresponding to its served MS. This means that the BS knows the channel coefficients from the MS to all other BSs in the network. This is not a very restrictive assumption as the MSs keep track of the channel coefficients of the neighboring BSs to perform tasks like the handoff.

With these assumptions, a decentralized algorithm for~\eqref{eq:statcl} can be obtained using the alternating direction method of multipliers (ADMM)~\cite{schizas,bazerque2011group,boyd2011distributed}. First, it is noted that~\eqref{eq:statcl} can be equivalently written as
\begin{align}
\min_{\bbW} \sum_{b=1}^{N_B} \left\{ \| \bar \bbI(:,b) - \bbW \bar \bbH(:,b) \|_F^2
+ \frac{\lambda}{N_B} \sum_{(c,c'): c \neq c'} s(\ccalB_c,\ccalB_{c'}) \right\} \label{eq:statcl2}
\end{align}
where $\bbM(:,b)$ denotes the $b$-th column of matrix $\bbM$. Then, a key step in deriving a decentralized solution is to introduce local auxiliary variables that decouple the problem across different BSs. Specifically, consider $\bar \bbW := \{\bbW_b\}_{b \in \ccalB}$ and $\bbU = [\{\bbU_b\}_{b \in \ccalB}, \{\bbU_e\}_{e \in \ccalE_c}]$. Then, \eqref{eq:statcl2} can be re-written as
\begin{align}
\min_{\bar \bbW,\bbU} & \sum_{b=1}^{N_B} \  \left[ \|\bar \bbI(:,b) - \bbU_b \bar \bbH(:,b)\|_F^2 + \frac{\lambda}{N_B} \sum_{(c,c'):c \neq c'} s_b(\ccalB_c,\ccalB_{c'}) \right] \label{eq:obj_dist_lasso} \\
\textrm{subject to } & \bbW_b = \bbU_b, \quad \forall b \in \ccalB   \label{eq:first_constraint} \\
& \bbW_b = \bbU_{(b,b')} = \bbW_{b'}, \quad \forall (b,b') \in \ccalE_c \label{eq:second_constraint}
\end{align}
where $s_b(\ccalB_c,\ccalB_{c'}) := \sum_{\bar b \in \ccalB_c} \sum_{\bar b' \in \ccalB_{c'}} \tilde w_{b,\bar b \bar b'}$ and $\{\tilde w_{b,\bar b \bar b'}\}$ are defined in the same way as $\{\tilde w_{\bar b \bar b'}\}$ are generated, but from $\bbW_b$, not $\bbW$. The idea is that each BS $b$ keeps a local copy $\bbW_b$ of $\bbW$, and \eqref{eq:second_constraint} enforces that the local copies of the neighboring BSs coincide. Thus, under the assumption that $\ccalG_c$ is connected,   $\bbW_b$ for all $b \in \ccalB$ are identical across the network. Also, at each BS $b$, $\bbU_b$ and $\bbW_b$ are constrained to be the same by~\eqref{eq:first_constraint}. Using separate $\bbW_b$ and $\bbU_b$ leads to decoupling of the MSE and the graph cut terms in \eqref{eq:obj_dist_lasso}, yielding closed-form local update rules.

To apply the ADMM technique, the Lagrangian is formed using the Lagrange multipliers $\bbV := [\{\bbV_b\}_{b \in \ccalB}, \{\bbV_e, \bar \bbV_e\}_{e \in \ccalE_c}]$ as
\begin{align}
\ccalL_\rho (\bar \bbW,\bbU,  \bbV) &:= \sum_{b=1}^{N_B} \left[\|\bar \bbI(:,b) - \bbU_b \bar \bbH(:,b)\|_F^2 + \frac{\lambda}{N_B} \sum_{(c,c'):c \neq c'} s_b(\ccalB_c,\ccalB_{c'}) \right] \nonumber\\
& + \sum_{b=1}^{N_B} \rho \left[ \langle \bbV_b, \bbW_b - \bbU_b \rangle + \frac{1}{2} \|\bbW_b - \bbU_b\|_F^2 \right] \nonumber  \\
& + \sum_{(b,b') \in \ccalE_c} \rho \left[ \langle \bbV_{(b,b')}, \bbW_b - \bbU_{(b,b')} \rangle + \frac{1}{2} \|\bbW_b - \bbU_{(b,b')} \|_F^2 \right. \nonumber\\
& \qquad\qquad + \left. \langle \bar \bbV_{(b,b')}, \bbW_{b'} - \bbU_{(b,b')} \rangle + \frac{1}{2} \|\bbW_{b'} - \bbU_{(b,b')} \|_F^2 \right]
\end{align}
where  $\rho > 0$ is a constant parameter and inner-product $\langle \bbM_1, \bbM_2 \rangle$ for complex matrices $\bbM_1$ and $\bbM_2$ is defined as $\langle \mathrm{Re}\{\bbM_1\},\mathrm{Re}\{\bbM_2\} \rangle + \langle \mathrm{Im}\{\bbM_1\},\mathrm{Im}\{\bbM_2\} \rangle$ with $\mathrm{Re}\{\cdot\}$ and $\mathrm{Im}\{\cdot\}$ denoting the real and the imaginary parts, respectively.

The ADMM consists of three steps: the $\bbV$-update, the $\bar \bbW$-update and the $\bbU$-update. The update for the Lagrange multipliers $\bbV$ can be done as
\begin{align}
&\bbV_b^{(k+1)} = \bbV_b^{(k)} + [\bbW_b^{(k)} - \bbU_b^{(k)}], \quad \forall b \label{eq:V_upd1}\\
&\bbV_{(b,b')}^{(k+1)} = \bbV_{(b,b')}^{(k)} + [\bbW_b^{(k)} - \bbU_{(b,b')}^{(k)}],\quad \forall (b,b') \in \ccalE_c \label{eq:V_upd2}\\
&\bar \bbV_{(b,b')}^{(k+1)} = \bar \bbV_{(b,b')}^{(k)} + [\bbW_{b'}^{(k)} - \bbU_{(b,b')}^{(k)}],\quad \forall (b,b') \in \ccalE_c \label{eq:V_upd3}
\end{align}
where superscript $^{(k)}$ signifies that the quantity corresponds to the $k$-th iteration.

\begin{table}[t]
\centering
\begin{tabular}{|l|}
\hline
Input: $\lambda$, $\rho$, clusters $\{\ccalB_c\}_{c=1}^{N_C}$\\
Output: $\{\bbW_b\}_{b=1}^{N_B}$\\
1: Initialize $\{\bbW_b^{(0)}\}_{b \in \ccalB}$, $\{\bbV_b^{(0)}\}_{b \in \ccalB}$, $\{\bbV_e^{(0)}\}_{e \in \ccalE_c}$, and $\{\bbU_b^{(0)}\}_{b \in \ccalB}$ as  zero matrices.\\
2: Repeat for $k = 0, 1, ...$ at each BS $b \in \ccalB$\\
3: \qquad Transmit $\bbW_b^{(k)}$ to neighbors.\\
4: \qquad Update $\bbV_b^{(k+1)}$ and $\bbV_e^{(k+1)}$ via \eqref{eq:V_upd1} and \eqref{eq:V_upd2_new}.\\
5: \qquad Update $\bbW_b^{(k+1)}$ via \eqref{eq:W_upd_new}.\\
6: \qquad Update $\bbU_b^{(k+1)}$ via \eqref{eq:U_upd1}.\\
7: \qquad $k \leftarrow k + 1$\\
8: Until convergence\\
9: Output $\{\bbW_b^{(k)}\}$.\\
\hline
\end{tabular}
\caption{Decentralized algorithm for static clustered MCP. }
\label{tab:alg-dist-fixed} \vspace{-10mm}
\end{table}

The update for $\bar \bbW$ is given by $\bar \bbW^{(k+1)} = \arg \min_{\bar \bbW} \ccalL_\rho(\bar \bbW, \bbU^{(k)}, \bbV^{(k+1)})$, which can be decomposed into individual BSs as
\begin{align}
\bbW_b^{(k+1)} &= \arg \min_{\bbW_b} \left[ \frac{\lambda}{N_B} \sum_{(c,c'):c \neq c'} s_b(\ccalB_c,\ccalB_{c'}) + \frac{\rho}{2} \|\bbW_b - \bbU_b^{(k)} + \bbV_b^{(k+1)} \|_F^2 \right.\nonumber\\
& \left. +  \frac{\rho}{2} \sum_{b': (b,b') \in \ccalE_c} \| \bbW_b - \bbU_{(b,b')}^{(k)} + \bbV_{(b,b')}^{(k+1)} \|_F^2 + \frac{\rho}{2} \sum_{b': (b',b) \in \ccalE_c} \| \bbW_b - \bbU_{(b',b)}^{(k)} + \bar \bbV_{(b',b)}^{(k+1)} \|_F^2 \right] \label{eq:W_upd}
\end{align}
which is essentially a group Lasso problem. The update for $\bbU$ is done similarly as
\begin{align}
\bbU^{(k+1)} = \arg \min_{\bbU} \ccalL_\rho (\bar \bbW^{(k+1)},\bbU, \bbV^{(k+1)})
\end{align}
which is equivalent to
\begin{align}
\bbU_b^{(k+1)} &= \arg \min_{\bbU_b} \|\bar \bbI(:,b) - \bbU_b \bar \bbH(:,b)\|_F^2 + \frac{\rho}{2} \|\bbW_b^{(k+1)} - \bbU_b + \bbV_b^{(k+1)}\|_F^2, \quad \forall b \in \ccalB \label{eq:U_upd1}\\
\bbU_{(b,b')}^{(k+1)} &= \arg \min_{\bbU_{(b,b')}} \frac{\rho}{2} \left( \|\bbW_b^{(k+1)} - \bbU_{(b,b')} + \bbV_{(b,b')}^{(k+1)}\|_F^2 + \|\bbW_{b'}^{(k+1)} - \bbU_{(b,b')} + \bar \bbV_{(b,b')}^{(k+1)} \|_F^2 \right) \nonumber\\
&= \frac{1}{2} \left( \bbW_b^{(k+1)} + \bbW_{b'}^{(k+1)} + \bbV_{(b,b')}^{(k+1)} + \bar \bbV_{(b,b')}^{(k+1)} \right), \quad \forall (b,b') \in \ccalE_c. \label{eq:U_upd2}
\end{align}
Now, it is observed that by summing up \eqref{eq:V_upd2}--\eqref{eq:V_upd3} and plugging in \eqref{eq:U_upd2}, one obtains
\begin{align}
\bbV_{(b,b')}^{(k+1)} + \bar \bbV_{(b,b')}^{(k+1)} = \bbV_{(b,b')}^{(k)} + \bar \bbV_{(b,b')}^{(k)} + \bbW_b^{(k)} + \bbW_{b'}^{(k)} - 2\bbU_{(b,b')}^{(k)} = {\bf 0}
\end{align}
which can simplify the update equations. Specifically, \eqref{eq:V_upd2} becomes
\begin{align}
\bbV_{(b,b')}^{(k+1)} = \bbV_{(b,b')}^{(k)} + \frac{1}{2} \left(\bbW_b^{(k)} - \bbW_{b'}^{(k)} \right),\quad \forall (b,b') \in \ccalE_c \label{eq:V_upd2_new}
\end{align}
and \eqref{eq:W_upd} becomes
\begin{align}
\bbW_b^{(k+1)} &= \arg \min_{\bbW_b} \left[ \frac{\lambda}{N_B} \sum_{(c,c'):c \neq c'} s_b(\ccalB_c,\ccalB_{c'}) + \frac{\rho}{2} \|\bbW_b - \bbU_b^{(k)} + \bbV_b^{(k+1)} \|_F^2 \right.\nonumber\\
& \qquad\qquad +  \frac{\rho}{2} \sum_{b': (b,b') \in \ccalE_c} \| \bbW_b - \frac{1}{2} \left(\bbW_b^{(k)} + \bbW_{b'}^{(k)} \right) + \bbV_{(b,b')}^{(k+1)} \|_F^2  \nonumber\\
& \left. \qquad\qquad + \frac{\rho}{2} \sum_{b': (b',b) \in \ccalE_c} \| \bbW_b - \frac{1}{2} \left(\bbW_b^{(k)} + \bbW_{b'}^{(k)} \right) - \bbV_{(b',b)}^{(k+1)} \|_F^2 \right], \quad \forall b \in \ccalB. \label{eq:W_upd_new}
\end{align}
Thus, the updates are necessary only for $\{\bbV_b\}_{b \in \ccalB}$, $\{\bbV_e\}_{e \in \ccalE_c}$, $\{\bbW_b\}_{b \in \ccalB}$, and $\{\bbU_b\}_{b \in \ccalB}$, via \eqref{eq:V_upd1}, \eqref{eq:V_upd2_new}, \eqref{eq:W_upd_new}, and \eqref{eq:U_upd1}, respectively. 

The overall algorithm is summarized in Table~\ref{tab:alg-dist-fixed}. In the begining of each iteration, all BSs transmit their latest copies of $\bbW_{b}^{(k)}$ to their one-hop neighbors. Then, each BS can locally carry out the updates in \eqref{eq:V_upd1} and \eqref{eq:V_upd2_new} to obtain $\bbV_{b}^{(k+1)}$ and $\bbV_{(b,b')}^{(k+1)}$, respectively. Note that $\bbV_{(b,b')}^{(k+1)}$ may be stored at both BSs $b$ and $b'$, which facilitates the subsequent update via \eqref{eq:W_upd_new}. After that, each BS can again locally update $\bbU_b^{(k+1)}$ using \eqref{eq:U_upd1}. In summary, all updates can be performed locally at individual BSs through exchanging $\bbW_b^{(k)}$ only with the one-hop neighbors in $\ccalG_c$. It can be shown that with $\rho > 0$, the iterates $\{\bbW_b^{(k)}\}_{b \in \ccalB}$ all converge to the solution to~\eqref{eq:statcl}~\cite{BeT89}.

\subsection{Decentralized Implementation for Dynamically Clustered MCP}
In case of dynamic clustering, solving~\eqref{eq:prob-rcut} in a decentralized fashion is of interest. As in the centralized implementation in Table~\ref{tab:alg-dynamic}, alternating minimization is employed, where $\bbW$ and $\{\ccalB_c\}$ are updated sequentially until no improvement in the cost can be made. The update for $\bbW$ while keeping clusters $\{\ccalB_c\}$ fixed is essentially an instantiation of static clustered MCP, which can be solved in a decentralized way via the algorithm in Table~\ref{tab:alg-dist-fixed}, with the only modification that the graph cut penalty $\lambda N_B^{-1} \sum_{(c,c'):c \neq c'} s_b(\ccalB_c,\ccalB_{c'})$ in~\eqref{eq:W_upd_new} is replaced by that of the ratio cut as $\lambda N_B^{-1} \sum_{c=1}^{N_C} s_b(\ccalB_c,\overline{\ccalB}_c)/|\ccalB_c|$.

After the execution of the algorithm in Table~\ref{tab:alg-dist-fixed}, all BSs have the same local copies of $\bbW$. Therefore, the BSs can simply use this $\bbW$ to perform spectral clustering to obtain the clusters $\{\ccalB_c\}_{c=1}^{N_C}$. However, this wastes computational resources since the same clustering problem is solved at all BSs. An alternative is to perform decentralized spectral clustering, as explained next.



Spectral clustering involves two basic steps. First, one needs to compute the eigenvectors corresponding to $N_C$ smallest eigenvalues of $\tilde \bbL$. Subsequently, the k-means algorithm must be executed on the rows of the matrix containing the $N_C$ eigenvectors as columns. Therefore, these two steps should be implemented in a decentralized fashion, as outlined in the next subsections. 

\begin{table}[t]
\centering
\begin{tabular}{|l|}
\hline
Input: $\bbA$ and $N_C$\\
Output: $\bbPhi := [\bbq_1 , \cdots, \bbq_{N_B} ]^T$\\
1: For each BS $b \in \ccalB$, randomly  initialize $\bbq_b^{(0)T} \in \mathbb{R}^{N_C}$, set $k = 0$, and run:\\
2: Repeat\\
3: \quad Update $\bbpsi_b^{(k+1)T} = \sum_{b' \in \ccalN_b} a_{bb'} \bbq_{b'}^{(k)T}$ \\
4: \quad Form $\bbK_b := \bbpsi_b^{(k+1)} \bbpsi_b^{(k+1)T}$ and perform consensus averaging to obtain $\bbK := \sum_{b=1}^{N_B} \bbK_b$.\\
5: \quad Compute Cholesky factorization on $\bbK  = \bbR^T \bbR$.\\
6: \quad Compute $\bbq_b^{(k+1)T} = \bbpsi_b^{(k+1)T} \bbR^{-1}$.\\
7: \quad $k \leftarrow k + 1$\\
8: Until convergence\\
9: Set $\bbq_b = \bbq_b^{(k)}$ for all $b \in \ccalB$.\\
10: For $N_c$ largest absolute values of eigenvalues, set $|\lambda_i| = \sqrt{K_{ii}},\ i =1,2,\ldots,N_C$.\\
\hline
\end{tabular}
\caption{Decentralized computation of eigenvectors of $\bbA$~\cite{Kempe:2004}.}
\label{tab:dist-e-vec} \vspace{-10mm}
\end{table}

\subsubsection{Decentralized Eigenvector Computation}
\label{sssec:dec_evec}

The algorithm in \cite{Kempe:2004} is an instance of the orthogonal iteration (OI), which is a generalization of the power method. The centralized OI would start with a random initial matrix $\bbPhi \in \mathbb{R}^{N_B \times N_C}$ and repeat the steps $\bbPsi \leftarrow \bbA \bbPhi$ and $\bbPhi \leftarrow \textrm{orthonormalize}(\bbPsi)$ until convergence. The resulting $\bbPhi$ matrix is the eigenvector matrix for $\bbA$.

In the decentralized OI, assuming that the $b$-th BS is equipped with the $b$-th row of $\bbA$, the $b$-th BS calculates only the $b$-th row of $\bbPhi$ through localized message passing. Let $\ccalN_b := \{b': a_{bb'} \neq 0,\ b' \in \ccalB\}$ be the set of neighbors of BS $b$ (including $b$ itself) based on affinity matrix $\bbA$. Then, the $b$-th BS starts with a randomly initialized nonzero vector $\bbq_b \in \mathbb{R}^{N_C}$, and forms $\bbpsi_b := \sum_{b' \in \ccalN_b} a_{bb'} \bbq_{b'}$, where $\{\bbq_{b'}\}_{b' \in \ccalN_b, b' \neq b}$ are collected from the one-hop neighbors. To orthonormalize $\bbPsi := [\bbpsi_1,\bbpsi_2,\ldots,\bbpsi_{N_B}]^T$ in a decentralized fashion, a QR decomposition is performed on $\bbPsi = \bbQ \bbR$. For this, the individual BSs form $\bbK_b := \bbpsi_b \bbpsi_b^T$ and run a consensus averaging iteration to obtain $\bbK := \sum_{b=1}^{N_B} \bbK_b = \bbPsi^T \bbPsi = \bbR^T \bbR \in \mathbb{R}^{N_C \times N_C}$. The consensus averaging can be performed by sharing information only among the one-hop neighbors in the connected communication graph. Then, $\bbR$ can be obtained at each BS via Cholesky factorization of $\bbK$. Subsequently, the $b$-th row of the orthonormalized version of $\bbPsi$ can be obtained by $\bbq_b^T = \bbpsi_b^T \bbR^{-1}$. Upon convergence, $\bbPsi = \bbA \bbPhi = \bbPhi \bbLambda$, where $\bbLambda$ is the diagonal matrix with the eigenvalues on the diagonal. Thus, $\bbK = \bbPsi^T \bbPsi = \bbLambda^2$. Therefore, the eigenvalue magnitudes are also obtained as the square roots of the diagonal entries in $\bbK$. The final algorithm is listed in Table~\ref{tab:dist-e-vec}.

Since our clustering algorithm requires the eigenvectors $\tilde \bbL$, corresponding to the $N_C$ smallest eigenvalues, a two-step procedure is used. First, the largest eigenvalue magnitude $|\lambda_1|$ is obtained via the algorithm in Table~\ref{tab:dist-e-vec} with $\bbA := \tilde \bbL$ and $N_C = 1$. Then, the same algorithm is executed again with $\bbA := (|\lambda_1| + \epsilon) \bbI - \tilde \bbL$, where $\epsilon$ is a small positive number, and $N_C$ set as the desired number of clusters. The last choice of $\bbA$ ensures that it is positive definite with the eigenvalue order is reversed from $\tilde \bbL$ as desired.

\subsubsection{Decentralized k-Means} \label{sssec:dec_kmeans}
To finally obtain the desired clusters $\{\ccalB_c\}_{c=1}^{N_C}$, the k-means algorithm must be performed on $\{\bbq_b\}$. A decentralized k-means algorithm developed in~\cite{FCG11jstsp} employs an ADMM procedure, and does not require the exchange of raw data among the agents, but only the local estimates of the centroids in the one-hop neighborhood. In our setup, each BS possesses data vector $\bbq_b$ from the decentralized eigendecomposition, and can readily execute this algorithm to obtain the cluster assignments. The decentralized $k$-means converges to a locally optimal solution~\cite{FCG11jstsp}.

The overall algorithm is the same as Table~\ref{tab:alg-dynamic}, with line 6 substituted by the procedure explained in Sec.~\ref{sssec:dec_evec}, line 7 by Sec.~\ref{sssec:dec_kmeans}, and line 8 by the algorithm in Table~\ref{tab:alg-dist-fixed} (with the ratio cut penalty).

\begin{figure}[t]
\centering
\centerline{\epsfig{figure=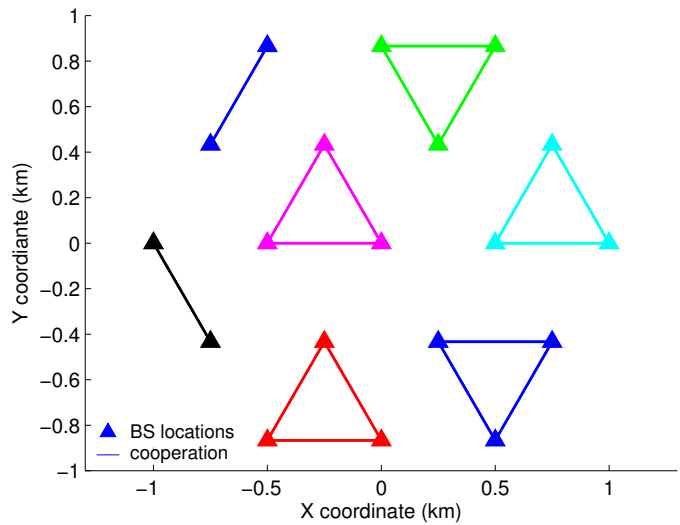,width=4in}}
\vspace{-5mm}
\caption{A network of 19 BSs used for numerical experiments. Locations of BSs are indicated by solid triangles. The BSs belonging to a cluster are depicted in the same color and connected by solid lines. The clusters shown here were used for experiments of static clustered MCP.}
\label{fig:clusters} \vspace{-3mm}
\end{figure}

\begin{remark} The communication overhead of the proposed algorithms is commensurate to how often the algorithms are run, which, in turn, should depend on the coherence time of the channels. Focusing on the MSs with slow channel variation, and also initializing the iterative algorithms with the iterates from the most recent round, the impact of the feedback overhead can be minimized. 
\end{remark}

\section{Numerical Tests}
\label{sec:test}
To test the proposed algorithms, a simple network comprising $19$  BSs with $500$~m cell radius is considered. The locations of the BSs are shown in Fig.~\ref{fig:clusters} as triangles. The single antenna MSs were dropped uniformly in the coverage area. The channels between the MSs and the BSs are composite of the path-loss, log-normal shadowing with standard deviation $8$~dB, and Rayleigh small-scale fading. The path-loss followed the 3GPP urban model given as $\mathrm{PL}(\mathrm{dB}) = 148.1 + 37.6 \log_{10} d$, where $d$ is the distance between the BS and the MS in~km. A path-loss exponent of $\alpha = 3.76$ was used. The MSs were assigned to the BSs with highest long-term channel gain (that is, the channel gain with the small-scale fading averaged out). When not otherwise noted, single-antenna BSs are assumed. In all the experiments, the number MSs served by each BS is equal to the number of antennas at the BS. Thus, the total number of MSs is equal to $N_A$.

\subsection{Distributed Cooperation}

\begin{figure}
\centering
\centerline{\epsfig{figure=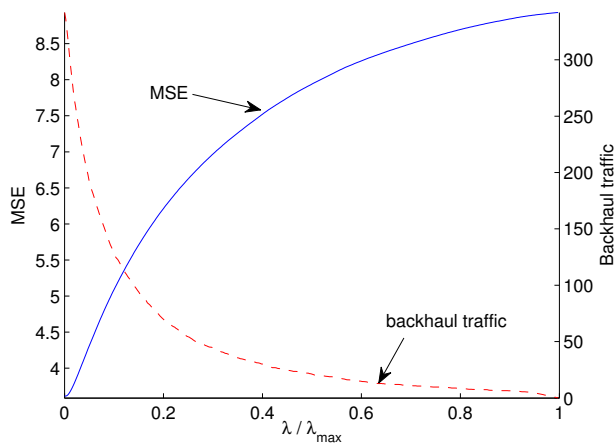,width=3.5in}}
\vspace{-8mm}
\caption{MSE incurred by distributed cooperation in the network shown in Fig. \ref{fig:clusters}. The number of antennas $A = 1$ was used.} \vspace{-5mm}
\label{fig:mse}
\end{figure}

\begin{figure}
\centering
\vspace{-5mm}
\centerline{\epsfig{figure=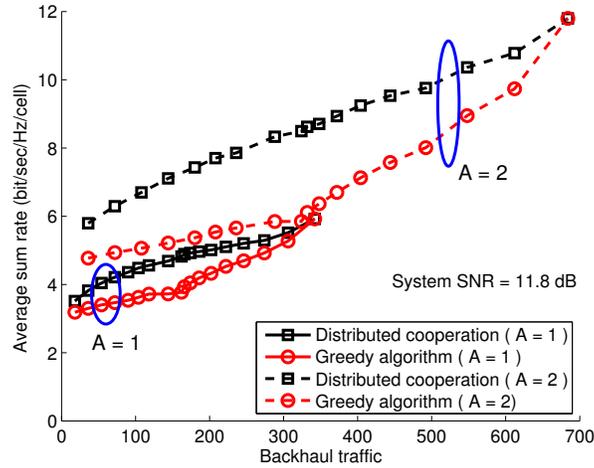,width=3.5in}}
\vspace{-9mm}
\caption{Average sum rates in distributed cooperation at various backhaul traffic levels. The curves with solid lines represent the case of single-antenna BSs $(A = 1)$ and the ones with dashed lines correspond to $A = 2$. The total number of MSs is the same as the total number of antennas. The proposed distributed cooperation outperforms the greedy algorithm in both cases.}
\label{fig:rate-vs-traffic} \vspace{-7mm}
\end{figure}

\begin{figure}
\centering
\centerline{\epsfig{figure=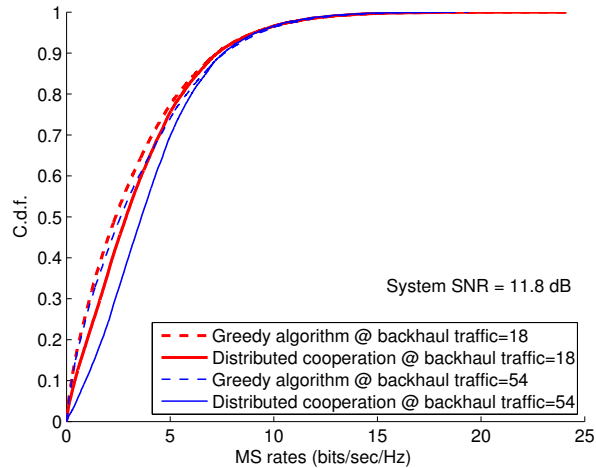,width=3.5in}}
\vspace{-8mm}
\caption{C.d.f.'s of MS rates at two values of backhaul traffic. Distributed cooperation increases the system fairness. }
\label{fig:cdf-distr} \vspace{-9mm}
\end{figure}

\begin{figure}
\centering
\centerline{\epsfig{figure=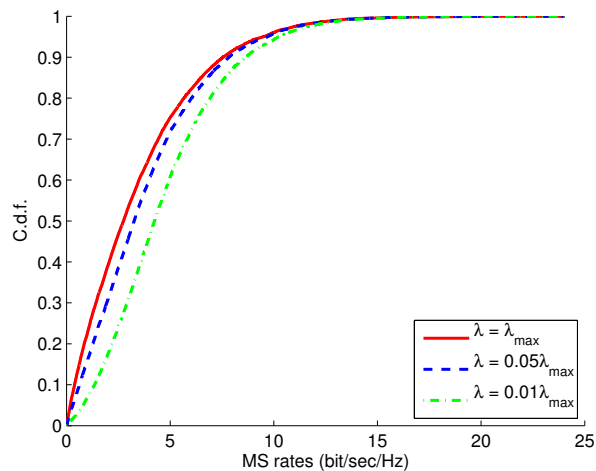,width=3.5in}}
\vspace{-9mm}
\caption{C.d.f.'s of MS rates at different backhaul traffic levels. Static clustered cooperation with the clusters shown in Fig.~\ref{fig:clusters} was used. The system fairness is improved as more inter-cluster backhaul traffic is allowed.}
\label{fig:cdf-static} \vspace{-5mm}
\end{figure}

The plots for MSE vs. backhaul traffic in distributed cooperation for 19 cell network of single antenna ($A=1$) BSs and 19 MSs is shown in Fig.~\ref{fig:mse}.  The system signal-to-noise-power-ratio~(SNR), defined as the average SNR at which the serving BS receives the signals from the MS located at the cell edge without accounting for inter-cell interference, was $6.2$~dB.  Parameter~$\lambmax$ denotes the value of $\lambda$ that yields sparse equalizer matrix with all off-diagonal entries as zero. Thus, $\lambda = \lambmax$ represents no cooperation, whereas $\lambda = 0$, full cooperation among all $19$ BSs. It can be seen that even with a small amount of backhaul traffic, significant reduction in MSE can be achieved. For instance, it can be seen that at $\lambda \approx 0.3 \lambmax$, only about $12\%$ of the backhaul traffic necessary for full cooperation (or about $40$ units) is incurred, but this achieves an MSE of about $7.0$. Compared to the MSE achievable by full cooperation at $3.6$, and that with no cooperation at $8.9$, this already captures about $36\%$ of the entire feasible reduction.
 

The performance of the proposed algorithm is compared to that of a greedy clustering algorithm in~\cite{PGH08}. The greedy clustering algorithm can be summarized as follows. Given a cluster size (the number of BSs per cluster), the greedy scheme picks the BS that yields the largest increase in the sum-rate as a member of the cluster sequentially. Once a cluster is formed, the algorithm moves on to the next cluster until all BSs belong to clusters. No inter-cluster backhauling is considered in the greedy scheme in~\cite{PGH08}. However, it is assumed that full cooperation is performed inside clusters. Fig.~\ref{fig:rate-vs-traffic} depicts the achieved average per-cell rates of the proposed distributed cooperation and the greedy scheme as functions of the total backhaul traffic incurred in the network, at the system SNR $11.8$~dB. Again the backhaul traffic in both the cases was equal to the $\ell_0$ norm of corresponding $\tilde \bbW$ matrices. The curves with solid lines represent the case of single-antenna BSs $(A = 1)$ and the ones with dashed lines correspond to $A = 2$. The total number of MSs is equal to the total number of antennas. In both cases, it is clearly seen that the proposed scheme achieves the trade-offs better than the greedy algorithm, and the gap between the two increases with the larger number of antenna elements.

Fig.~\ref{fig:cdf-distr} shows the cumulative distribution function (c.d.f.) of the rates of the individual MSs achieved through distributed cooperation of single-antenna BSs. Two sets of curves are plotted. One set was obtained at the backhaul traffic amount of $18$ units, and the other set of $54$ units. In each set, the solid curve represents the c.d.f. due to the proposed algorithm, and the dashed one corresponds to that of the greedy scheme presented for comparison. For greedy clustering, the cluster sizes of $2$ and $4$ yield the backhaul traffic of $18$ and $54$, respectively. For distributed cooperation, the value of $\lambda$ was adjusted to yield the same amounts of backhaul traffic. It can be observed that the cell-edge users enjoy far better rates under the proposed distributed cooperation at the same backhaul traffic level. 

\begin{figure}
\centering
\begin{minipage}{0.45\textwidth}
\centerline{\epsfig{figure=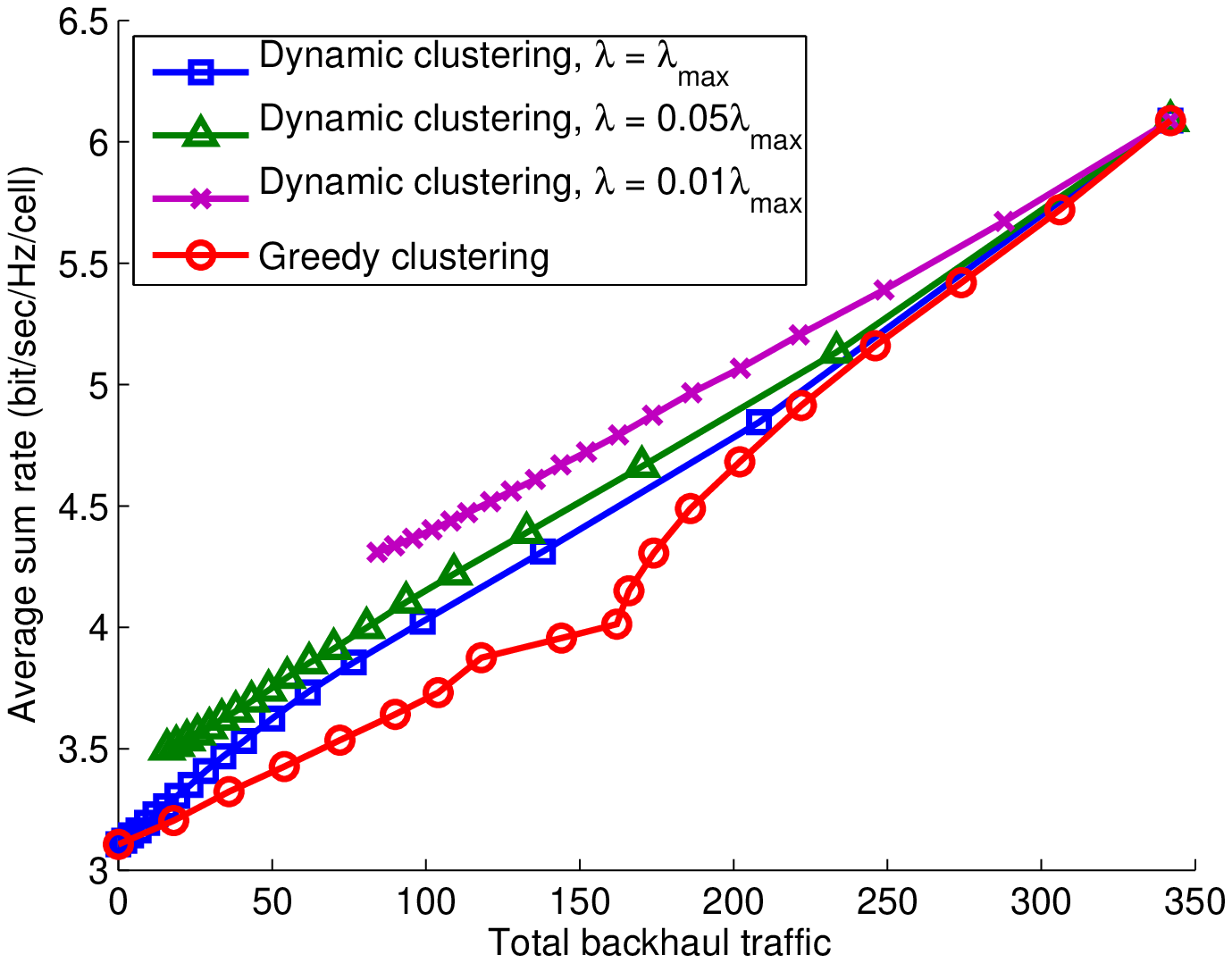,width=3.5in}}

{\small (a) Full distributed cooperation within clusters}
\end{minipage} \hfill
\begin{minipage}{0.45\textwidth}
\centerline{\epsfig{figure=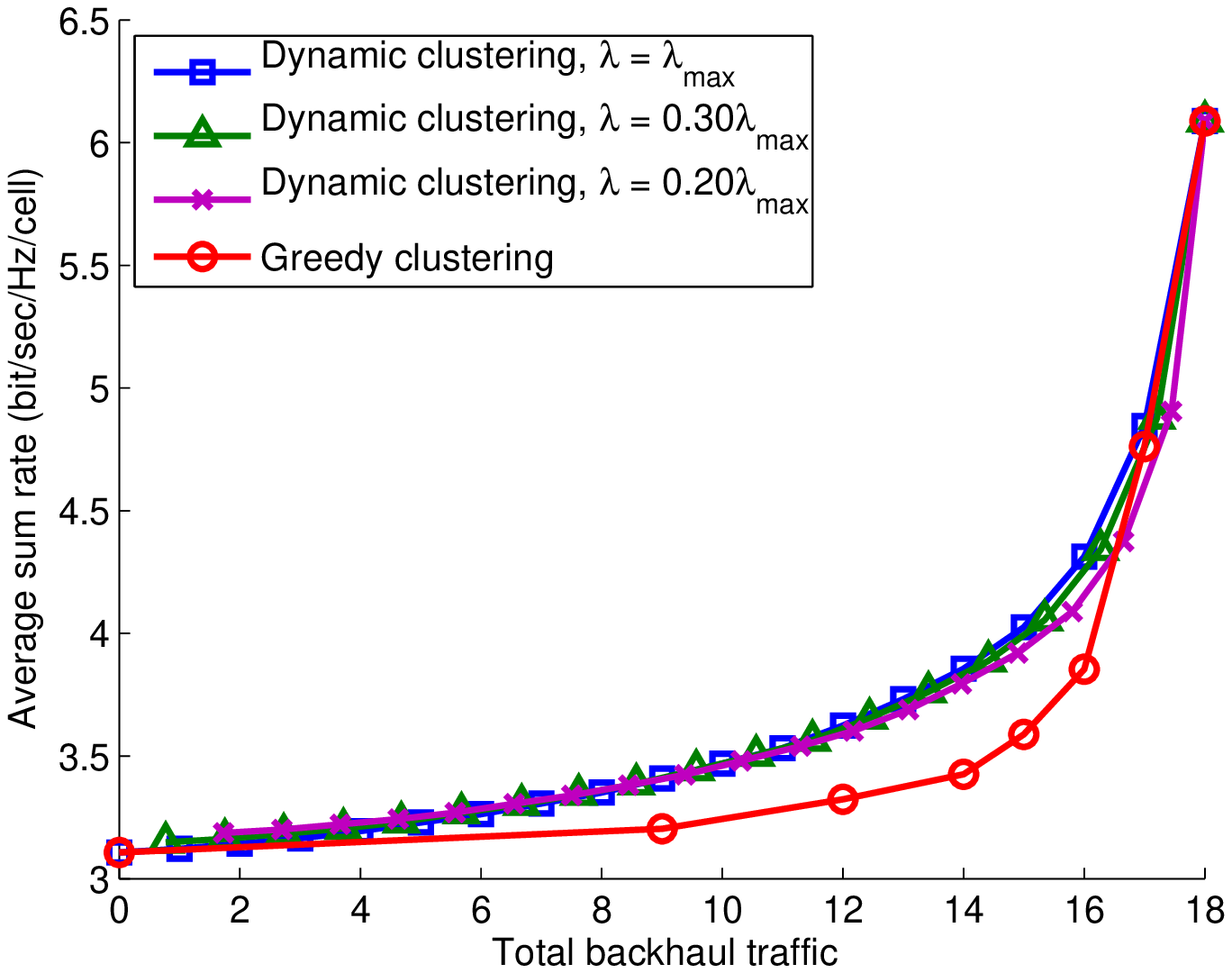,width=3.5in}}

{\small (b) Centralized cooperation within clusters}
\end{minipage}

\caption{Average sum rates in dynamic clustered cooperation at various backhaul traffic levels. Panel (a) is for the case where full distributed cooperation takes place within the clusters. Panel (b) corresponds to the case where centralized cooperation takes place at the cluster heads. Dynamic clustered cooperation outperforms greedy clustering in both cases.}
\label{fig:rate-dynamic} \vspace{-5mm}
\end{figure}

\begin{figure}
\begin{minipage}{0.45\textwidth}
\centering
\epsfig{figure=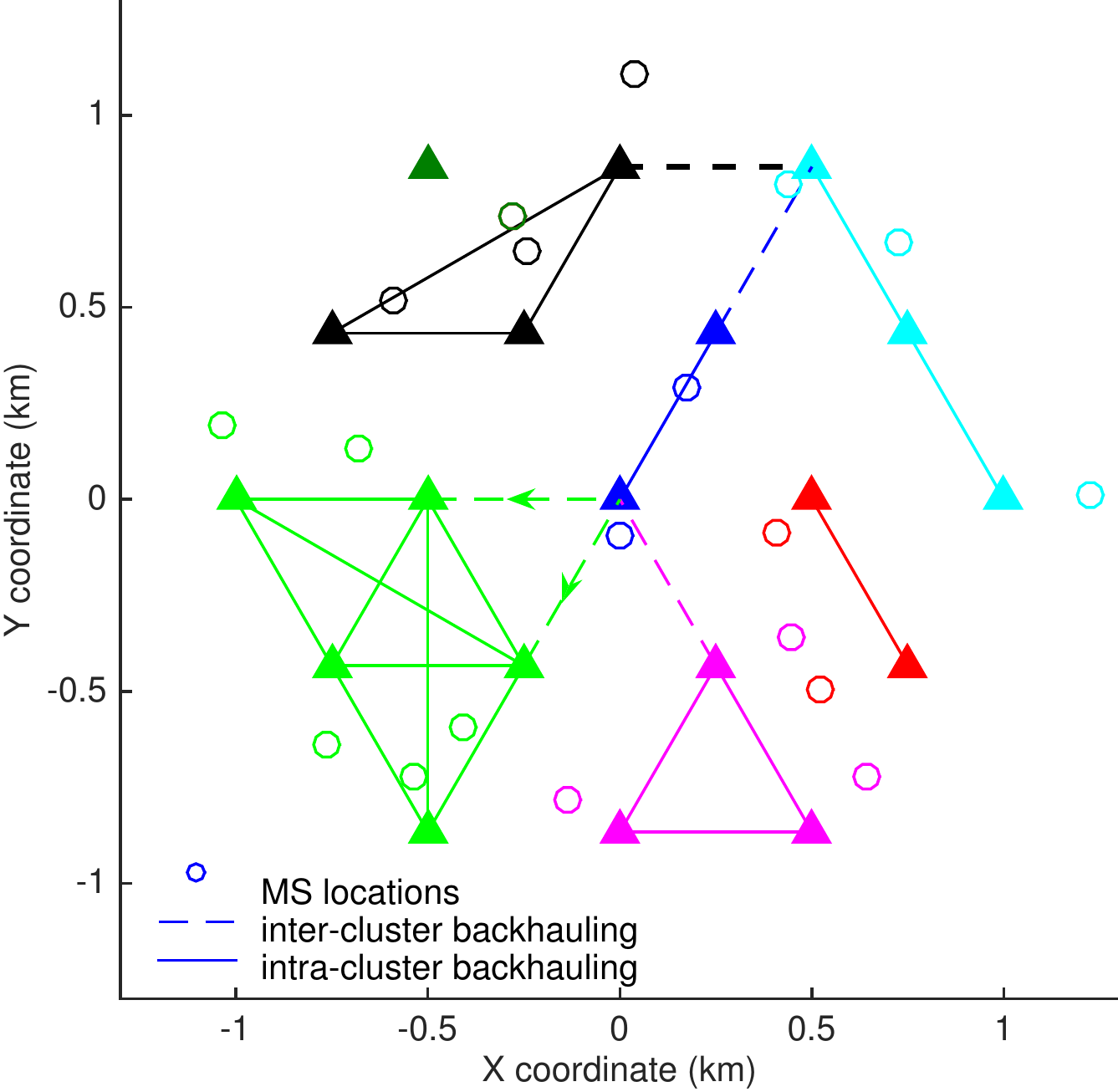,width=3in}

{\small (a) no shadowing, system SNR = $11.2$ dB}
\end{minipage} \hfill
\begin{minipage}{0.45\textwidth}
\centering
\epsfig{figure=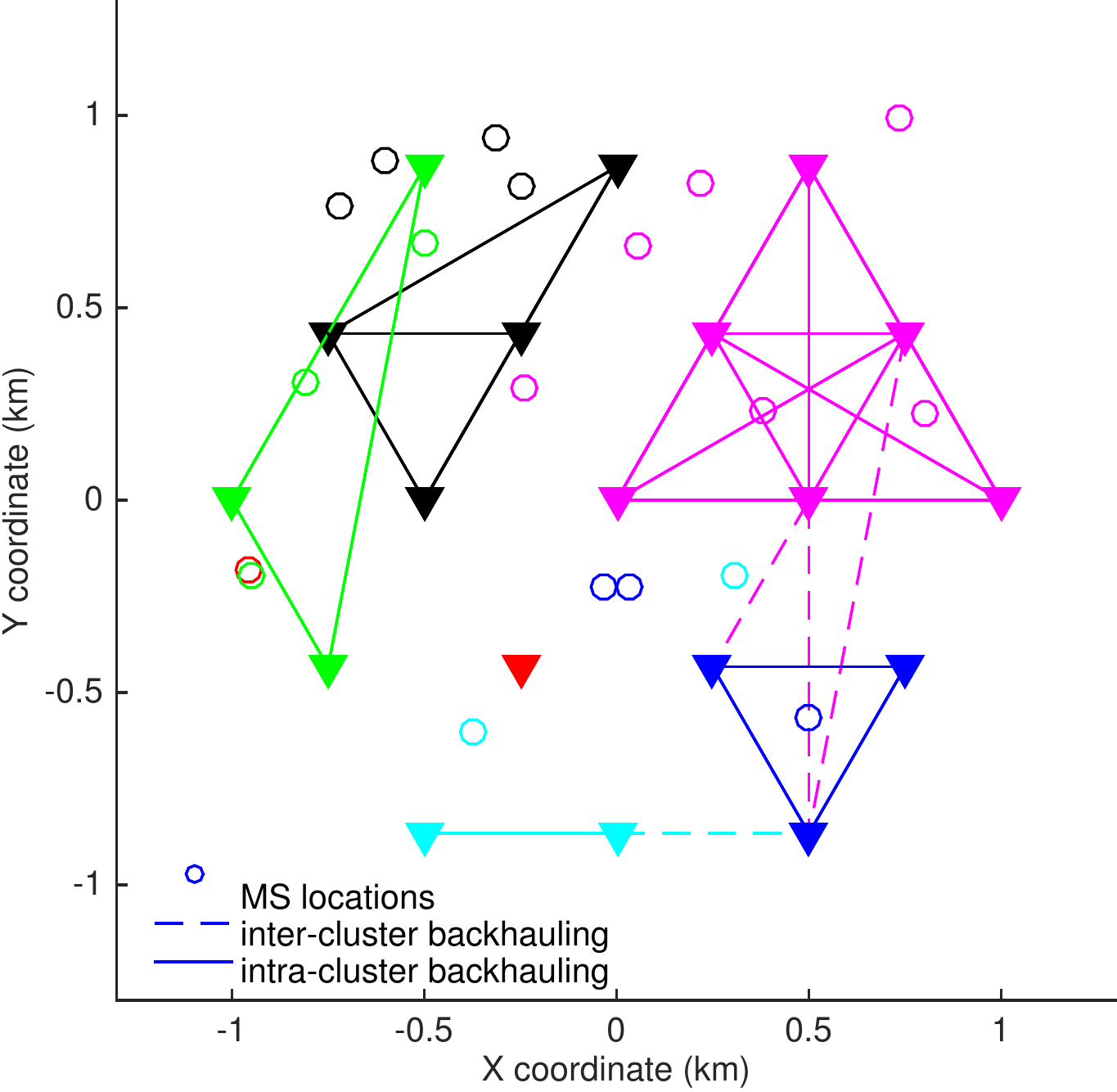,width=3in}

{\small (b) with shadowing, system SNR = $16.2$ dB}
\end{minipage}
\caption{Clusters obtained by dynamic clustering algorithm for the network of $19$ single-antenna BSs/MSs shown in Fig.~\ref{fig:clusters}. Each MS is served by the BS with highest channel gain. The clusters shown are for $\lambda = 0.1 \lambmax$.}	\label{fig:clexample}
\vspace{-3mm}
\end{figure}

\subsection{Clustered Cooperation}

First, the performance of the static clustered cooperation is verified. A seven-cluster partitioning as shown in Fig.~\ref{fig:clusters} was employed. Single-antenna BSs were employed and full cooperation was used within clusters. The number of MSs in the system is $19$. Due to the fixed clustering, the MSs at the cluster edges may experience excessive interference. Fig.~\ref{fig:cdf-static} depicts the c.d.f. of the MS rates with varying degrees of inter-cluster cooperation. The curve for $\lambda = \lambmax$ represents the performance without any inter-cluster cooperation. The average inter-cluster backhaul amounts due to $\lambda = 0.05 \lambmax$ and $\lambda = 0.01\lambmax$ turned out to be $28.5$ and $124$, respectively. It can be seen that the rates of the cluster-edge mobiles are improved significantly through inter-cluster cooperation. 

To assess performance of the dynamic clustered cooperation, the average per-cell rates for different backhaul traffic amounts are plotted in Fig.~\ref{fig:rate-dynamic} at the system SNR equal to $11.8$~dB.  The algorithm in Table~\ref{tab:alg-dynamic} was used with the initial equalizer $\bbW_0$ set to the LMMSE equalizer $\bbW_{\mathrm{lmmse}}$. Although we did not formally prove the convergence of the algorithm, it always converged within 2-3 iterations in our experiments. For comparison, the curve from the greedy clustering case is shown again. The markers in the greedy clustering curve correspond to different cluster sizes, which range from $1$ to $19$. Similarly, the markers in the dynamic clustering curves represent a variable number of clusters~$N_C$, which also ranges from $1$ to $19$. The effect of inter-cluster cooperation was examined by varying the value of $\lambda$, where $\lambda = \lambmax$ again signifies no inter-cluster cooperation. The backhaul traffic includes both the inter- and the intra-cluster backhauling. Fig.~\ref{fig:rate-dynamic}(a) was plotted by modeling the intra-cluster traffic for a cluster $\cV_c$ as $|\cV_c|(|\cV_c|-1)$ units, which represents the case where distributed cooperation occurs inside the clusters. On the other hand, the intra-cluster traffic per cluster $\cV_c$ was assumed equal to $|\cV_c|-1$ units in Fig.~\ref{fig:rate-dynamic}(b), which corresponds to the situation where a BS inside each cluster is elected as the cluster head for centralized cooperation inside the cluster, with all the other cluster members feeding their samples to the cluster head. It can be seen that even without inter-cluster cooperation, the dynamic clustering algorithm outperforms the greedy scheme under both models of intra-cluster backhaul traffic. The performance is seen to be further improved by having inter-cluster cooperation, especially at low backhaul traffic, which is the regime of practical interest. However, this additional gain seems to be limited, in particular when central processing per cluster is assumed, as the spectral clustering-based dynamic BS partitioning itself has already absorbed the major portion of the MCP gain.

\begin{figure}
\centering
\centerline{\epsfig{figure=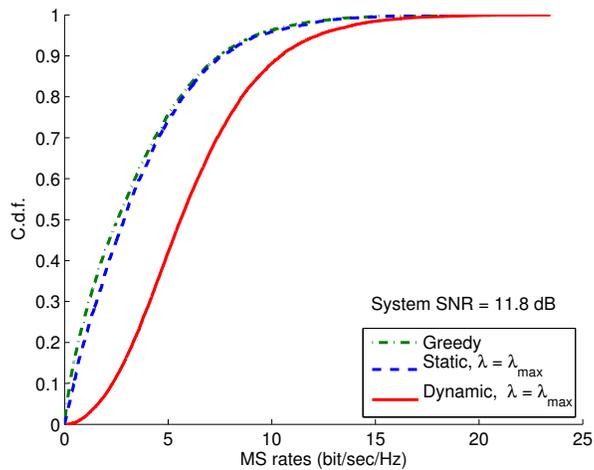,width=3.5in}}
\vspace{-5mm}
\caption{C.d.f.'s of MS rates with dynamic clustered cooperation. The network in Fig. \ref{fig:clusters} with $A=1$ was used. Full cooperation was used within the clusters and no inter-cluster cooperation was allowed. Dynamic clustered cooperation performs the best, followed by the static clustered and the greedy scheme.}
\vspace{-5mm}
\label{fig:cdf-dynamic}
\end{figure}

\begin{figure}
\centering
\begin{tabular}{cc}
\includegraphics[width=0.4\linewidth]{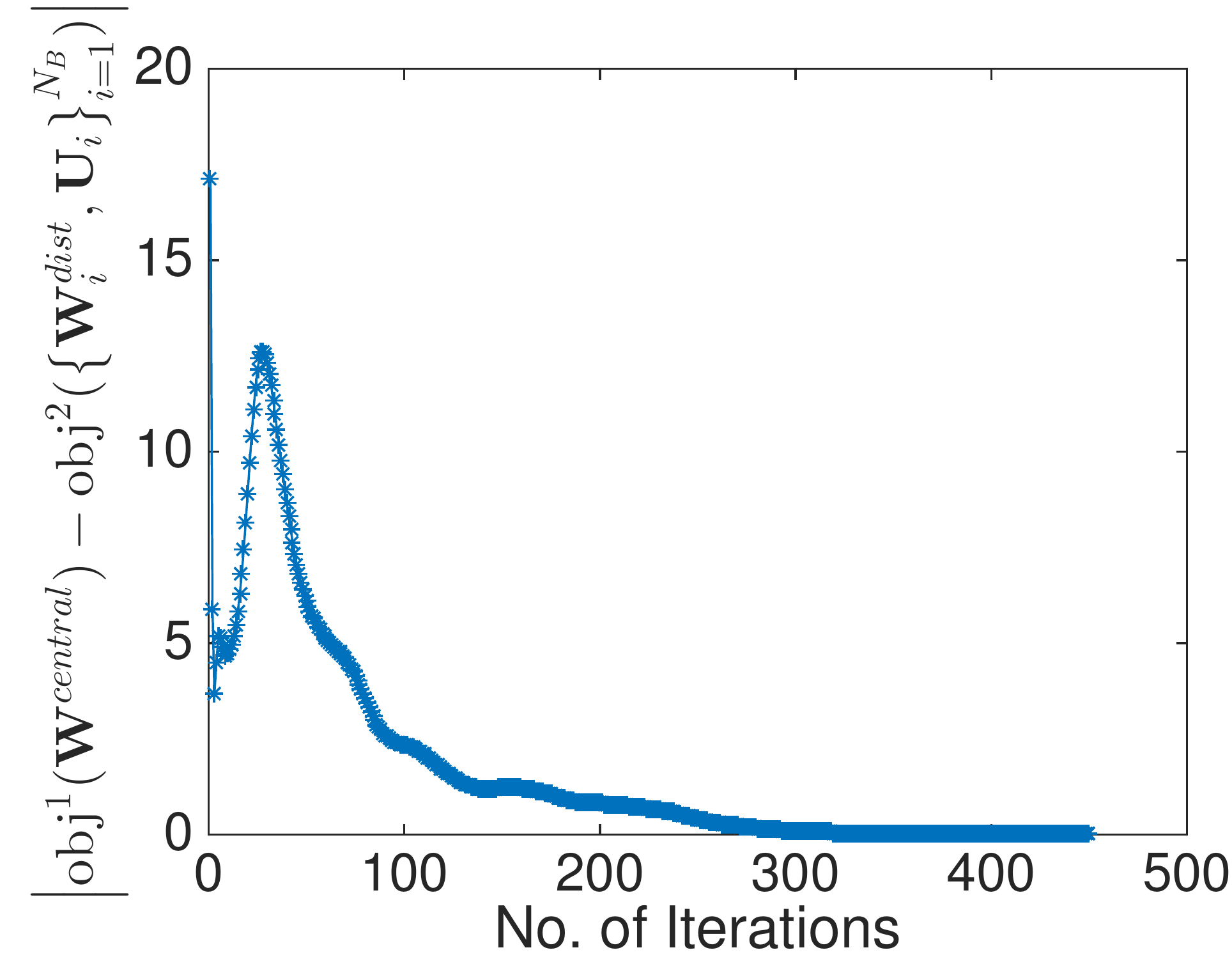}
\includegraphics[width=0.4\linewidth]{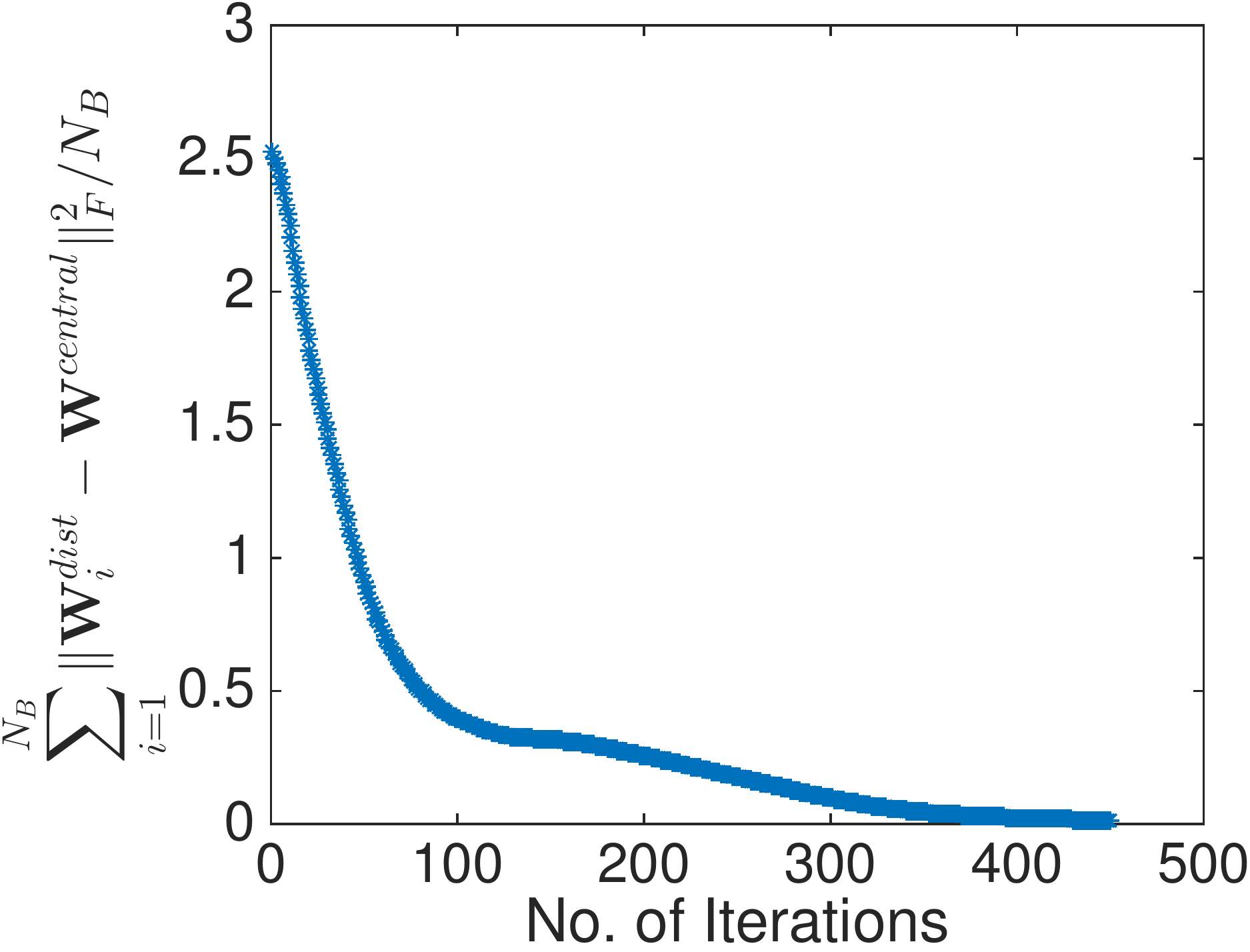}
\end{tabular}
\caption{Convergence of the decentralized algorithm for static clustered MCP.}
\label{fig:cong_dist_lasso}
\end{figure}

\begin{figure}
\centering
\begin{tabular}{cc}
	\includegraphics[width=0.4 \linewidth]{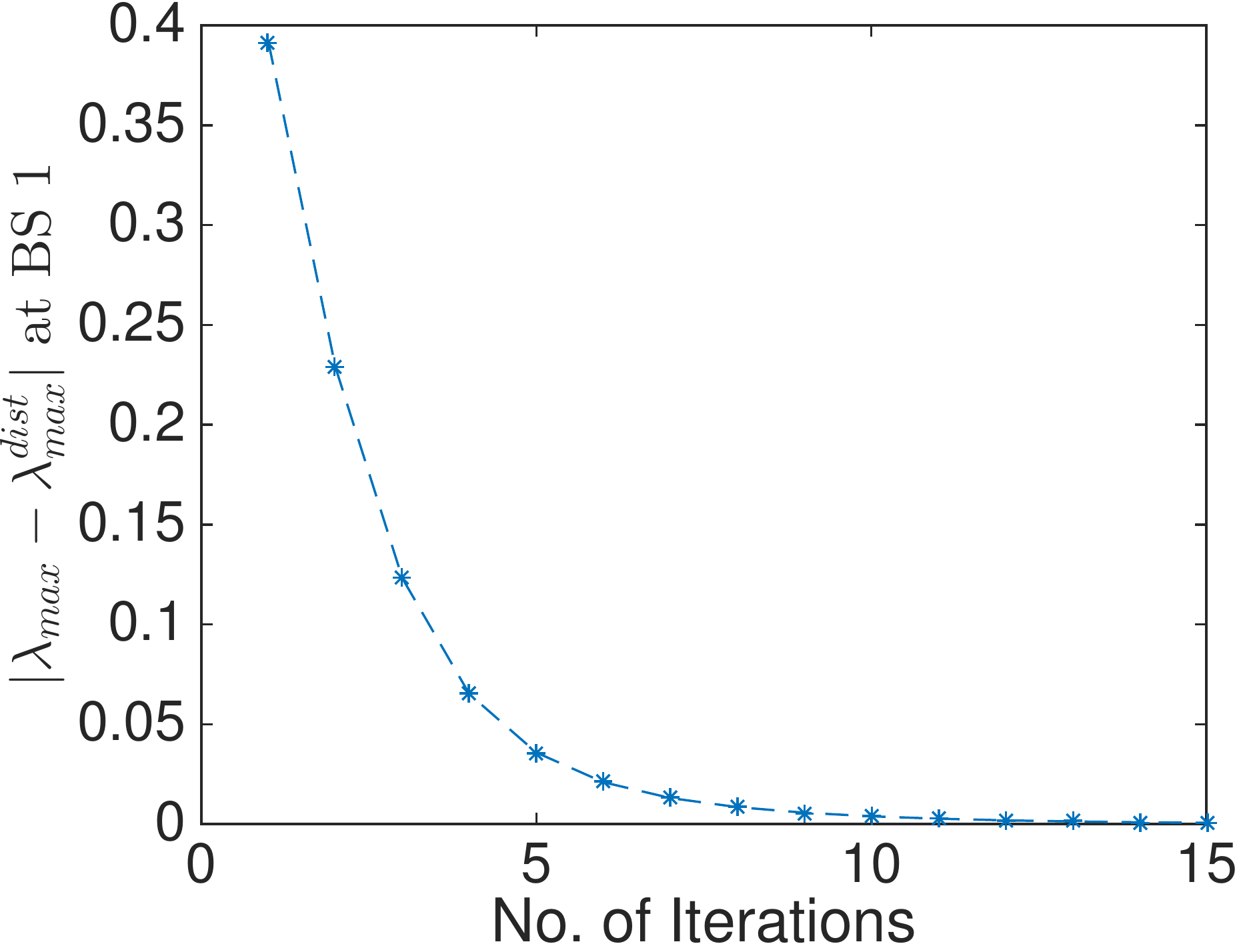}
	\includegraphics[width=0.4 \linewidth]{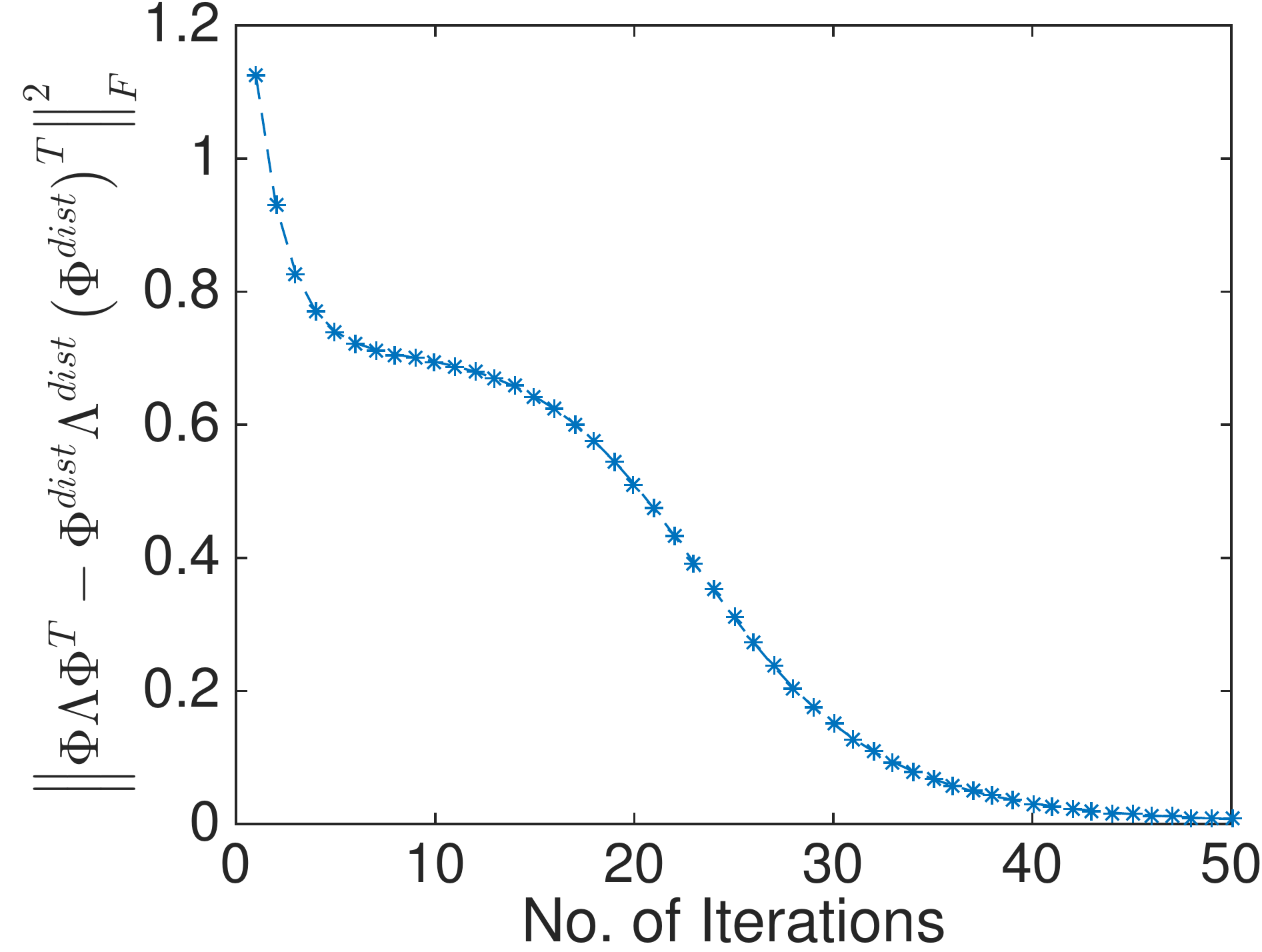}
\end{tabular}
\caption{Convergence of decentralized eigenvector computation.}\label{fig:cong_dist_ev}
\vspace{-3mm}
\end{figure}

Instances of cooperating clusters formed by the dynamic clustering algorithm are depicted in Fig.~\ref{fig:clexample}. The solid lines represent the (bi-directional) intra-cluster cooperation links, and the dashed arrows denote the inter-cluster cooperation, obtained with $\lambda = 0.1 \lambmax$. To check that the formed clusters make intuitive sense, the shadowing and the small-scale fading were suppressed when generating {Fig.~\ref{fig:clexample}(a)}. Thus, each MS is served by the closest BS, and the cluster formation is solely dictated by the geometry of the BS and the MS locations. The MS locations are marked by the circles in Fig.~\ref{fig:clexample}. It can be observed that the clusters are mostly formed in such a way to protect the cell-edge users. Fig.~\ref{fig:clexample}(b) presents the case with shadow fading and the system SNR equal to $16.2$ dB. Each MS is associated with the BS with highest channel gain, which might not be the closest BS necessarily. The color of each MS is matched to that of the cluster by which the MS is served.

The c.d.f. curve of the MS rates for dynamic clustered cooperation with $N_C = 7$ and $\lambda = \lambmax$ is shown in Fig.~\ref{fig:cdf-dynamic}. For comparison, the curves corresponding to the greedy scheme and the static clustered cooperation with the clusters in Fig.~\ref{fig:clusters} are also provided. The figure underlines the clear improvement in fairness using the proposed dynamic clustering even without inter-cluster cooperation.

\subsection{Convergence of Decentralized Implementation}
\label{subsec:decen_group_lasso}

To illustrate the convergence of the decentralized algorithm for static clustered MCP in Table~\ref{tab:alg-dist-fixed}, the left panel in Fig.~\ref{fig:cong_dist_lasso} depicts the evolution of the difference between the objective in \eqref{eq:statcl2} for the centralized optimization and the objective \eqref{eq:obj_dist_lasso} for the decentralized optimization is plotted. The clusters in Fig.~\ref{fig:clusters} is again used with $\lambda = 0.4 \lambmax$ and $\rho = 0.1$. The communication graph for the decentralized algorithm was chosen so that each BS can communicate with its nearest neighbors. The plot shows that the decentralized objective quickly converges to that of the centralized algorithm. In the right panel of Fig.~\ref{fig:cong_dist_lasso}, the convergence of the iterates of the equalizer matrix is shown. It can be observed that the decentralized solution reaches global consensus toward the centralized solution.

The convergence of the decentralized eigenvector computation via the algorithm in Table~\ref{tab:dist-e-vec} is illustrated in Fig.~\ref{fig:cong_dist_ev}. The sparse equalizer obtained in Fig.~\ref{fig:cong_dist_lasso} was used to obtain $\tilde \bbL$, and the largest eigenvalue was first computed using algorithm in Table~\ref{tab:dist-e-vec}. The left panel of Fig.~\ref{fig:cong_dist_ev} shows the evolution of the difference between the centrally {\color{green}computed} eigenvalue and the local copy at BS~$1$. It can be seen that the local copy converges quickly to the central solution. After this, $N_C = 9$ top eigenvectors of $\bbA = (\lambda_{max}^{dist}+ 0.1)\bbI - \tilde \bbL$ are computed. The right panel of Fig.~\ref{fig:cong_dist_ev} depicts the convergence of $\bbPhi^{dist} \bbLambda^{dist} \left(\bbPhi^{dist}\right)^T$, where $\bbPhi^{dist}$ contains the $N_C$ top eigenvectors from the distributed algorithm, and $\bbLambda^{dist}$ is the diagonal matrix with $N_C$ top eigenvalues of $\bbA$. Again, the convergence is seen to be quite fast.

\section{Conclusions}
\label{sec:conc}
Backhaul-constrained MCP was considered for cellular uplinks. Exploiting recent compressive sensing techniques, a reduced-backhaul linear equalizer was obtained. In the case of clustered cooperation, sparsity was promoted on inter-cluster feedback. For the dynamic clustering set-up, a \emph{joint} cluster formation and equalization problem was formulated, and an iterative algorithm based on spectral clustering was developed. Decentralized implementations of the latter were introduced, and validated with simulations. Numerical tests also demonstrated that significant reduction in MSE is possible at small increase in backhaul traffic. 

\bibliographystyle{IEEEtran}
\bibliography{mcp}


\end{document}